\documentclass[runningheads]{llncs}

\usepackage{amsmath}
\usepackage{graphicx}
\usepackage{amsfonts, amssymb}
\usepackage{xspace}
\usepackage{algorithm}
\usepackage{algpseudocode}
\usepackage{ltl}
\usepackage{latexsym, stmaryrd, xspace}
\usepackage{paralist}
\usepackage{todonotes}
\usepackage{listings}
\usepackage{array}%
\usepackage{color, colortbl}
\usepackage{tikz}
\usetikzlibrary{arrows, arrows.meta,decorations.pathreplacing,calligraphy}
\tikzset{arw/.style={>={Triangle[length=3mm,width=3mm]},line width=2mm,draw=gray}}
\usepackage{tabularx}
\usepackage{graphicx,lipsum, wrapfig}
\usepackage{caption}
\usepackage{framed}
\usepackage{subcaption}
\usepackage{multirow}
\usepackage{makecell}
\usepackage{booktabs}

\usepackage{colortbl}

\usepackage[misc]{ifsym}

\usepackage{hyperref}
\hypersetup{%
	colorlinks=true,           
	allcolors=blue!70!black,   
	pdfstartview=Fit,          
	breaklinks=true,
	pdfauthor={T.-H. Hsu, B. Bonakdarpour, and B. Finkbeiner},
	pdftitle={Bounded Model Checking of Asynchronous Hyperproperties}
}

\usepackage{tikz}
\usetikzlibrary{calc}

\tikzset{
	>=stealth,
	possible world/.style={circle,draw,thick,align=center},
	real world/.style={double,circle,draw,thick,align=center}, 
}
\usetikzlibrary{
	automata,
	backgrounds,
	shapes, 
	arrows,
}

\pgfdeclarelayer{bg}    
\pgfdeclarelayer{fg}    
\pgfsetlayers{bg,main, fg}

\lstdefinestyle{CStyle}{
    backgroundcolor=\color{backgroundColour},   
    commentstyle=\color{mGreen},
    keywordstyle=\color{magenta},
    numberstyle=\tiny\color{mGray},
    stringstyle=\color{mPurple},
    basicstyle=\footnotesize,
    breakatwhitespace=false,         
    breaklines=true,                 
    captionpos=b,                    
    keepspaces=true,                 
    numbers=left,                    
    numbersep=2pt,                  
    showspaces=false,                
    showstringspaces=false,
    showtabs=false,                  
    tabsize=2,
    numberblanklines=false,
    escapeinside=||,
    numberfirstline=false,
firstnumber=1,
    language=C
}

\makeatletter
    \def\@citecolor{blue}%
    \def\@urlcolor{blue}%
    \def\@linkcolor{blue}%
 \def\UrlFont{\rmfamily}
 \def\orcidID#1{\smash{\href{http://orcid.org/#1}{\protect\raisebox{-1.25pt}{\protect\includegraphics{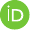}}}}}
 \makeatother

\newcommand{\LTL}{\textsf{\small LTL}\xspace}
\newcommand{\HyperLTL}{\textsf{\small HyperLTL}\xspace}

\newcommand{\AHLTL}{\textsf{\small A-HLTL}\xspace}

\newcommand{\Paths}{\ensuremath{\mathsf{Paths}}\xspace}
\newcommand{\Trajs}{\ensuremath{\mathsf{Trajs}}\xspace}
\newcommand{\Vars}{\Paths}
\newcommand{\Traces}{\ensuremath{\mathsf{Traces}}\xspace}

\newcommand{\AP}{\ensuremath{\mathsf{AP}}\xspace}

\newcommand{\PTR}{\ensuremath{\mathsf{PTR}}\xspace}

\newcommand{\MC}[2]{{\mbox{\sf \small MC\big[#1\big]}$^{#2}$\xspace}}

\newcommand{\code}[1]{\texttt{#1}}
\newcommand{\V}{\mathcal{V}}
\newcommand{\T}{\mathcal{T}}
\newcommand{\J}{\mathcal{J}}

\newcommand{\tru}{\mathtt{true}}
\newcommand{\fals}{\mathtt{false}}

\newcommand{\HyperQube}{\textsc{HyperQB}\xspace}
\newcommand{\MCHyper}{\textsc{MCHyper}\xspace}
\newcommand{ \defAs}{\triangleq}

\newcommand{\state}{s}
\newcommand{\trace}{\sigma}
\newcommand{\traj}{t}
\newcommand{\trvar}{\pi}
\newcommand{\tjvar}{\tau}
\newcommand{\trass}{\Pi}
\newcommand{\tjass}{\Gamma}
\newcommand{\tjall}{\ensuremath{\mathsf{TRJ}}\xspace}
\newcommand{\alphabet}{\mathrm{\Sigma}}
\newcommand{\naturals}{\mathbb{N}_0}

\newcommand{\krip}{\mathcal{K}}
\newcommand{\ktuple}{\langle S, s_\init, \trans, L \rangle}
\newcommand{\init}{\mathit{init}}

\newcommand{\Trace}{\mathsf{Traces}}
\newcommand{\States}{S}
\newcommand{\trans}{\delta}

\newcommand{\comp}[1]{\textsf{\small #1}}
\newcommand{\quant}{\mathbb{Q}}

\newcommand{\sync}{\mathsf{sync}}
\newcommand{\start}{\mathsf{start}}

\newcommand{\Atau}{\mathsf{A}}
\newcommand{\Etau}{\mathsf{E}}

\renewcommand{\Until}{\mathbin\mathcal{U}}
\newcommand{\F}{\LTLdiamond}

\newcommand{\Always}{\LTLsquare}
\newcommand{\Event}{\LTLdiamond} 
\newcommand{\Next}{\LTLcircle}

\newcommand{\true}{\mathit{true}}
\newcommand{\false}{\mathit{false}}
\newcommand{\into}{\ensuremath{\rightarrow}}
\renewcommand{\And}{\mathrel{\wedge}}
\newcommand{\Or}{\mathrel{\vee}}

\newcommand{\Iff}{\mathrel{\leftrightarrow}}

\newcommand{\Into}{\mathrel{\rightarrow}}
\newcommand{\IntoP}{\mathrel{\rightharpoonup}}
\newcommand{\partTo}{\IntoP}
\newcommand{\Nat}{\mathbb{N}}

\newcommand{\sem}[1]{\ensuremath{\llbracket #1\rrbracket}}
\newcommand{\QBF}[1]{\ensuremath{\llbracket #1 \rrbracket}}

\newcommand{\Pos}{\ensuremath{\textit{pos}}}
\newcommand{\Dom}{\ensuremath{\textit{Dom}}}
\newcommand{\Enc}{\ensuremath{\textit{enc}}}

\newcommand{\ie}{i.e.,\xspace}

\newcommand{\myend}{\mathsf{end}}

\newcommand{\obs}{\mathsf{obs}}

\newcommand{\kframe}{\mathcal{F}}

\newcommand{\DefinedAs}{\ensuremath{\,\stackrel{\text{\textup{def}}}{=}\,}}

\newcommand{\map}{\mathsf{map}}
\definecolor{mGreen}{rgb}{0,0.6,0}
\definecolor{mGray}{rgb}{0.5,0.5,0.5}
\definecolor{mPurple}{rgb}{0.58,0,0.82}
\definecolor{backgroundColour}{rgb}{0.95,0.95,0.92}

\makeatletter
\def\@citecolor{blue}%
\def\@urlcolor{blue}%
\def\@linkcolor{blue}%
\def\UrlFont{\rmfamily}
\def\orcidID#1{\smash{\href{http://orcid.org/#1}{\protect\raisebox{-1.25pt}{\protect\includegraphics{orcid_color.eps}}}}}
\makeatother

\newcommand{\PR}[1]{\ensuremath{\textit{#1}}}

\newcommand{\Halt}{\PR{halt}\xspace}
\newcommand{\Halted}{\PR{halted}\xspace}

\newcommand{\HOPT}{\mathit{hopt}}
\newcommand{\HPES}{\mathit{hpes}}

\newcommand{\HOmodels}{\models^{\HOPT}}
\newcommand{\HPmodels}{\models^{\HPES}}

\newcommand{\off}{\PR{off}\,}

\newcommand{\release}{\mathbin{\mathcal{R}}}

\newcommand{\ENDS}{\ensuremath{\PR{ends}\,}}
\newcommand{\STEP}{\ensuremath{\PR{step}\,}}
\newcommand{\MOVES}{\ensuremath{\PR{moves}\,}}
\newcommand{\STUTTERS}{\ensuremath{\PR{stutters}\,}}
\newcommand{\SETPOS}{\ensuremath{\PR{setpos}\,}}
\newcommand{\NOPOS}{\ensuremath{\PR{nopos}\,}}

\newcommand{\policy}{\mathsf{p}}

\newcommand{\propNI}{ \textsf{NI}}
\newcommand{\propNIndet}{ {\textsf{NI}_\textsf{nd}}  }
\newcommand{\propOD}{\textsf{OD}}
\newcommand{\propODndet}{{\textsf{OD}_\textsf{nd}} }
\newcommand{\opt}{\mathsf{SC}}
\newcommand{\optndet}{ {\textsf{SC}_\textsf{nd}}  }
\newcommand{\propSNI}{\textsf{SNI}}

\newcommand{\prophigh}{\textsf{h}}
\newcommand{\proplow}{\textsf{l}}

\newcommand{\propin}{\textsf{in}}
\newcommand{\propout}{\textsf{out}}

\newcommand{\progndet}{\texttt{ndet}}
\newcommand{\progacdb}{\texttt{ACDB}}
\newcommand{\progxy}{\texttt{ConcLeak}}
\newcommand{\progse}{\texttt{SpecExcu}}
\newcommand{\progdbe}{\texttt{DBE}}
\newcommand{\proglp}{\texttt{LP}}
\newcommand{\progeflp}{\texttt{EFLP}}
\newcommand{\progcta}{\texttt{CacheTA}}

\newcommand{\SAT}{\texttt{SAT}}
\newcommand{\UNSAT}{\texttt{UNSAT}}
\newcommand{\PPos}{\varphi_\Pos}

\definecolor{prettygreen}{rgb}{0.0, 0.5, 0.0}
\definecolor{prettyred}{rgb}{1.0, 0.13, 0.32}
\definecolor{prettyblue}{rgb}{0.0, 0.30, 1.0}
\newcommand{\truevalue}{\color{prettygreen}}
\newcommand{\falsevalue}{\color{black!55}}

\begin{document}

\newcommand{\Thanks}{This research has been partially
		supported by the United States NSF SaTC Award 2100989,
		by the Madrid Regional Gov. Project BLOQUES-CM
		(S2018/TCS-4339), by Project PRODIGY
		(TED2021-132464B-I00) funded by \linebreak
		MCIN/AEI/10.13039/501100011033/ and the EU
		NextGenerationEU/PRTR, by the German Research Foundation (DFG) as part of TRR 248 
		(389792660), and by the European Research Council (ERC) Grant HYPER (101055412)}

\title{Bounded Model Checking for \\ Asynchronous Hyperproperties
}

\makeatletter
\def\@citecolor{blue}%
\def\@urlcolor{blue}%
\def\@linkcolor{blue}%
\def\UrlFont{\rmfamily}
\def\orcidID#1{\smash{\href{http://orcid.org/#1}{\protect\raisebox{-1.25pt}{\protect\includegraphics{orcid_color.eps}}}}}
\makeatother

\author{Tzu-Han Hsu\inst{1}\orcidID{0000-0002-6277-2765}
	\and \Letter~Borzoo Bonakdarpour\inst{1}\orcidID{0000-0003-1800-5419}
	\and Bernd Finkbeiner\inst{2}\orcidID{0000-0002-4280-8441} 
	\and  \\ César Sánchez\inst{3}\orcidID{0000-0003-3927-4773}
}

\authorrunning{T.-H. Hsu et al.}

\institute{
	Michigan State University, East Lansing, MI, USA \email{\{tzuhan,borzoo\}@msu.edu}
	\and
CISPA Helmholtz Center for Information Security, Saarbrücken, Germany
\email{finkbeiner@cispa.de}
	\and
	IMDEA Software Institute, Madrid, Spain \email{cesar.sanchez@imdea.org}
}

\maketitle

	\begin{abstract}
    Many types of attacks on confidentiality stem from the
    nondeterministic nature of the environment that computer programs
    operate in (e.g., schedulers and asynchronous communication
    channels).
	In this paper, we focus on verification of confidentiality in
	nondeterministic environments by reasoning about {\em
	asynchronous hyperproperties}.
	First, we generalize the temporal logic \AHLTL to allow nested
	{\em trajectory} quantification, where a trajectory determines
	how different execution traces may advance and stutter.
	We propose a bounded model checking algorithm for \AHLTL based
	on QBF-solving for a fragment of the generalized \AHLTL and evaluate it by various case studies 
	on
	concurrent programs, scheduling attacks, compiler
	optimization, speculative execution, and cache timing attacks.
	We also rigorously analyze the complexity of model checking
	for different fragments of \AHLTL.
\end{abstract}


	
	


\section{Introduction}
\label{sec:intro}

\begin{wrapfigure}{r}{.26\textwidth}
	\vspace{-10mm}
	\scalebox{.83}{
	\input{figs/prog1}
}
	\vspace{-8mm}
	\caption{\code{T1} and \code{T2} leak the value of \code{h}.}
	\label{fig:prog1}
	\vspace*{-10mm}
\end{wrapfigure}
\subsubsection{Motivation.}
Consider the concurrent program~\cite{g07} shown in
Fig.~\ref{fig:prog1}, where \code{h} is a secret variable, and
\code{await} command is a conditional critical region.
%
This program should satisfy the following information-flow policy:

\vspace{-0.3em}
\begin{quote}
\emph{ ``Any sequences of observable outputs produced by an
	interleaving should be reproducible by some other interleaving
	for a different value of \code{h}''. }
\end{quote}
\vspace{-0.3em}
If this is the case, then an attacker cannot successfully guess the
value of \code{h} from the sequence of observable outputs of the
\code{print()} statements.
For example, Fig.~\ref{fig:v1good} shows how one can align two
interleavings of threads \code{T1} and \code{T2} with respect to the
observable sequence of outputs \code{`abcd'},
\begin{figure*}[b!]
	\hspace*{-1mm}
	\centering
	\scalebox{.67}{
		\begin{tikzpicture}
	\tikzstyle{eclp}=[draw=black, ellipse, minimum width=26pt, minimum height=22pt,
	align=center]
	
	\begin{scope}
		\node [eclp](a1) {---};
		
		\node [above=10pt] at (a1) {T1-2};
		\node [left=20pt] at (a1) {\code{h=0}};
		
		\node[eclp, left=150pt, right=30pt] (a2) at (a1) {---};
		\node [above=10pt] at (a2) {T1-3};
		
		\node[eclp, left=150pt, right=30pt] (a3) at (a2) {`\code{a}'};
		\node [above=10pt] at (a3) {T1-4};
		
		\node[eclp, left=150pt, right=30pt] (a4) at (a3) {---};
		\node [above=10pt] at (a4) {T1-5};
		
		\node[eclp, left=150pt, right=30pt] (a5) at (a4) {`\code{b}'};
		\node [above=10pt] at (a5) {T1-6};
		
		\node[eclp, left=150pt, right=30pt] (a6) at (a5) {`\code{c}'};
		\node [above=10pt] at (a6) {T2-11};
		
		\node[eclp, left=150pt, right=30pt] (a7) at (a6) {---};
		\node [above=10pt] at (a7) {T2-12};
		
		\node[eclp, left=150pt, right=30pt] (a8) at (a7) {---};
		\node [above=10pt] at (a8) {T1-7};
		
		\node[eclp, left=150pt, right=30pt] (a9) at (a8) {---};
		\node [above=10pt] at (a9) {T2-18};
		
		\node[eclp, left=150pt, right=30pt] (a10) at (a9) {`\code{d}'};
		\node [above=10pt] at (a10) {T2-19};
%
%
		\draw[->] (a1) -- (a2);
		\draw[->] (a2) -- (a3);
		\draw[->] (a3) -- (a4);
		\draw[->] (a4) -- (a5);
		\draw[->] (a5) -- (a6);
		\draw[->] (a6) -- (a7);
		\draw[->] (a7) -- (a8);
		\draw[->] (a8) -- (a9);
		\draw[->] (a9) -- (a10);
	\end{scope}	
	
	\begin{scope}[yshift=-1.3cm]
		\node [eclp](b1) {---};
		\node [below=10pt] at (b1) {T1-2};
		\node [left=20pt] at (b1) {\code{h=1}};
		
		\node[eclp, left=150pt, right=30pt] (b2) at (b1) {---};
		\node [below=10pt] at (b2) {T1-3};
		
		\node[eclp, left=150pt, right=30pt] (b3) at (b2) {`\code{a}'};
		\node [below=10pt] at (b3) {T1-4};
		
		\node[eclp, left=150pt, right=30pt] (b4) at (b3) {---};
		\node [below=10pt] at (b4) {T1-5};
		
		\node[eclp, left=150pt, right=30pt] (b5) at (b4) {`\code{b}'};
		\node [below=10pt] at (b5) {T1-6};
		
		\node[eclp, left=150pt, right=30pt] (b6) at (b5) {`\code{c}'};
		\node [below=10pt] at (b6) {T2-11};
		
		\node[eclp, left=150pt, right=30pt] (b7) at (b6) {---};
		\node [below=10pt] at (b7) {T2-12};
		
		\node[eclp, left=150pt, right=30pt] (b8) at (b7) {---};
		\node [below=10pt] at (b8) {T1-7};
		
		\node[eclp, left=150pt, right=30pt] (b9) at (b8) {---};
		\node [below=10pt] at (b9) {T2-13,14};
		
		\node[eclp, left=150pt, right=30pt] (b10) at (b9) {---};
		\node [below=10pt] at (b10) {T2-15};
		
		\node[eclp, left=150pt, right=30pt] (b11) at (b10) {---};
		\node [below=10pt] at (b11) {T2-16};

		\node[eclp, left=150pt, right=30pt] (b12) at (b11) {`\code{d'}};
\node [below=10pt] at (b12) {T2-19};


		\draw[->] (b1) -- (b2);
		\draw[->] (b2) -- (b3);
		\draw[->] (b3) -- (b4);
		\draw[->] (b4) -- (b5);
		\draw[->] (b5) -- (b6);
		\draw[->] (b6) -- (b7);
		\draw[->] (b7) -- (b8);
		\draw[->] (b8) -- (b9);
		\draw[->] (b9) -- (b10);
		\draw[->] (b10) -- (b11);
		\draw[->] (b11) -- (b12);
		
	\end{scope}
	
	\draw[-, dashed] (a1) -- (b1);
	\draw[-, dashed] (a2) -- (b2);
	\draw[-, dashed] (a3) -- (b3);
	\draw[-, dashed] (a4) -- (b4);
	\draw[-, dashed] (a5) -- (b5);
	\draw[-, dashed] (a6) -- (b6);
	\draw[-, dashed] (a7) -- (b7);
	\draw[-, dashed] (a8) -- (b8);
	\draw[-, dashed] (a9) -- (b9);
	\draw[-, dashed] (a9) -- (b10);
	\draw[-, dashed] (a9) -- (b11);
	\draw[-, dashed] (a10) -- (b12);

\end{tikzpicture}
	}
	\caption{Two secure interleavings for the program in Fig.~\ref{fig:prog1}}
	\label{fig:v1good}	
	\vspace*{-12mm}
\end{figure*}
given two different values of secret \code{h}. Let us call such an alignment a {\em trajectory} (illustrated 
by the sequence of dashed lines).
However, if thread \code{T1} holds the semaphore and executes the critical 
region as an atomic operation.
Then, output \code{‘acdb’} arising due to concurrent execution of threads \code{T1} and 
\code{T2} reveals the value of \code{h} as 0, as the same output cannot be reproduced when 
\code{h=1}.
Thus, the program in Fig.~\ref{fig:prog1} violates the above policy.

\begin{wrapfigure}{r}{.28\textwidth}
	\vspace*{-8mm}
	\scalebox{.83}{
		\input{figs/prog2}
	}
	\vspace*{-6mm}
	\caption{\code{T1} and \code{T2} receive inputs from 
	asynch. channels read by \code{T3} and \code{T4}.}
	\label{fig:prog2} 
	\vspace{-6mm}
\end{wrapfigure} 
The above policy is an example of a {\em hyperproperty}~\cite{cs10}; i.e., a 
set of sets of execution traces.
In addition to information-flow requirements, hyperproperties can
express other complex requirements such as linearizability~\cite{hw90}
and control conditions in cyber-physical systems such as robustness
and sensitivity.
%
%
The temporal logic \AHLTL~\cite{bcbfs21} can  express hyperproperties 
whose sets of traces advance at different speeds, allowing stuttering steps.
For example, the above policy can be expressed in \AHLTL by the following formula:
$
\varphi_\propNI = \forall \pi.\exists \pi'. \Etau \tau.
(\prophigh_{\pi, \tau} \neq \prophigh_{\pi', \tau})\wedge 
\Always(\obs_{\pi,\tau} = \obs_{\pi', 
	\tau}),
$
where $\obs$ denotes the output observations, meaning that for all executions (i.e., 
interleavings) $\pi$, there should exist another execution $\pi'$ 
and a trajectory $\tau$, such that $\pi$ and $\pi'$ start from different values of \code{h} and $\tau$ 
can align all the observations along $\pi$ and $\pi'$ (see Fig.~\ref{fig:v1good}).
\AHLTL can reason about {\em one} source of {\em nondeterminism} by the scheduler in the system that may 
lead to information leak.
Indeed, the model checking algorithms proposed in~\cite{bcbfs21} can discover the bug in the 
program in Fig.~\ref{fig:prog1}.

%

%
%

Now, consider a more complex version of the same program shown in Fig.~\ref{fig:prog2} inspired 
by modern programming languages such as \code{Go} and \code{P} that allow  
CSP-style concurrency.
Here, new threads \code{T3} and \code{T4} read the values of secret input \code{h} and public input 
\code{l} from two asynchronous channels,
%
%
rendering two different sources of nondeterminism: (1) the scheduler that results in 
different interleavings, and (2) data availability in the channels.
This, in turn, means formula $\varphi_\propNI$ no longer captures the following specification of the program, 
which should be:

\begin{quote}
	{\em
		``Any sequence of observable outputs produced by an interleaving should be reproducible by 
		some other interleaving such that for all alignments of public inputs, there exists an alignment of 
		the public outputs''.
	}
\end{quote}
Satisfaction of this policy (not expressible in \AHLTL as proposed 
in~\cite{bcbfs21})
prohibits an attacker from successfully determining the 
sequence of values of \code{h}.

%

\subsubsection{Contributions.}
%
In this paper, we strive for a general logic-based approach that enables 
model checking of a rich set of asynchronous hyperproperties. 
To this end, we concentrate on \AHLTL model checking for programs 
subject to multiple sources of nondeterminism.
Our first contribution is a generalization of \AHLTL that allows nested {\em trajectory} quantification.
For example, the above policy requires reasoning about two different trajectories that 
cannot be composed into one since their sources of nondeterminism are different.
%
%
%
%
This observation motivates the need for enriching \AHLTL with the tools to quantify over trajectories.
This generalization enables expressing policies such as follows:
%
$$
	\varphi_\propNIndet = \forall \pi.\exists \pi'. \Atau \tau. \Etau \tau'. 
	  \big( \F(\prophigh_{\pi, \tau} \neq \prophigh_{\pi', \tau}) \land  \
	\Always\ (\proplow_{\pi, \tau} = \proplow_{\pi', \tau}) \big) 
	\rightarrow  \Always (\obs_{\pi,\tau'} = \obs_{\pi', \tau'}),
$$
%
%
where $\Atau$ and $\Etau$ denote the universal (res., existential) trajectory quantifiers.

Our second contribution is a {\em bounded model checking} (BMC) 
algorithm for a fragment of the extended \AHLTL that allows an arbitrary 
number of trace quantifier alternations and up to one  trajectory quantifier 
alternation.
Following~\cite{hsb21}, we propose two bounded semantics (called {\em optimistic} and {\em 
pessimistic}) for \AHLTL based on the satisfaction of eventualities.
We introduce a reduction to the satisfiability problem for quantified 
Boolean formulas (QBF) and prove that our translation provides decision procedures for \AHLTL 
BMC for {\em terminating systems}, i.e., those whose Kripke structure is acyclic.
Our focus on terminating programs is due to the general undecidability of 
\AHLTL model checking~\cite{bcbfs21}.
As in the classic BMC for \LTL, the power of our technique is in hunting 
bugs that are often in the shallow parts of reachable states.

\begin{wraptable}{r}{.4\textwidth}
	\centering
	\vspace{-8mm}
	
	\renewcommand{\arraystretch}{1.7}
	\newcolumntype{K}[1]{>{\centering\arraybackslash}p{#1}}
	\scalebox{.55}{
	\begin{tabular}{|K{3.6cm} || K{4cm}| K{.8cm} |}
		\hline
		\multicolumn{3}{|c|}{\cellcolor{black!5} \bf Multiple Traces --  Single Trajectory}\\
		\hline\hline
		$\exists^+\Etau$ / $\forall^+(\Atau/\Etau) $ & 
		\multicolumn{2}{|c|}{\makecell{\comp{NL-complete} \\ 
		(\footnotesize Theorem~\ref{thm:altfree-single})}}\\	
		\hline
		$\big[\exists(\exists/\forall)^+(\Atau/\Etau)\big]^k$ & \comp{${\Sigma^p_{k}}$-complete} & 
		\multirow{2}{*}{\rotatebox{270}{\footnotesize Thm~\ref{thm:alternating-single}}}\\
		\cline{1-2}
		$\big[\forall(\exists/\forall)^+(\Etau/\Atau)\big]^k$ & \comp{${\Pi^p_{k}}$-complete} &\\
		\hline\hline
		\multicolumn{3}{|c|}{\cellcolor{black!5} \bf Multiple Traces --  Multiple Trajectories}\\
		\hline\hline
		$\big[\exists(\exists/\forall)^+(\Etau^+\Etau)\big]^k$ & 
		\comp{$\Sigma_{k+1}^p$-complete} & \multirow{2}{*}{\rotatebox{270}{\footnotesize 
		Thm~\ref{thm:alternating-multiple}}}\\
		\cline{1-2}
		$\big[\forall(\forall/\exists)^+(\Atau^+\Atau)\big]^k$ & 
\comp{$\Pi_{k+1}^p$-complete} &\\
		\hline
		$\big[\exists(\exists/\forall)^+\Atau^+\Etau^+\big]^k$ & 
		\comp{$\Sigma_{k+1}^p$-complete} & 
		\multirow{2}{*}{\rotatebox{270}{\footnotesize 
		Thm~\ref{thm:alternating-multiple-alternating}}}\\
		\cline{1-2}
		$\big[\forall(\forall/\exists)^+\Etau^+\Atau^+\big]^k$ & \comp{$\Pi_{k+1}^p$-complete} 
		& \\
\hline
\AHLTL &  \multicolumn{2}{|c|}{\comp{PSPACE}}  \\
\hline
	\end{tabular}
}

	\caption{\AHLTL model checking complexity for acyclic models.}
	\label{fig:table}
	\vspace{-8mm}
\end{wraptable}
Our third contribution is rigorous complexity analysis of \AHLTL model checking for terminating 
programs (see Table~\ref{fig:table}).
We show that for formulas with only one trajectory quantifier the complexity is aligned with that of  
classic synchronous semantics of \HyperLTL~\cite{clarkson14temporal}.
However, the complexity of  \AHLTL model checking with multiple trajectory quantifiers is one step 
higher than \HyperLTL model checking in the polynomial hierarchy.
An interesting observation here is that the complexity of model checking a formula with two 
existential trajectory quantifiers is one step higher than one with only one existential quantifier 
although the plurality of the quantifiers does not change. Generally 
speaking, \AHLTL model checking for terminating 
programs remains in \comp{PSPACE}.

Finally, we have implemented our BMC technique.
%
%
We evaluate our implementation on verification of four case studies: (1) information-flow 
security in concurrent programs, (2) information leak in speculative executions, (3) preservation of 
security in compiler optimization, and (4) cache-based timing attacks.
These case studies exhibit a proof of concept for the highly intricate nature 
of information-flow requirements and how our foundational theoretical 
results handle them.

\subsubsection{Related Work.}
%
The concept of hyperproperties is due to Clarkson and Schneider~\cite{cs10}.
\HyperLTL~\cite{clarkson14temporal} and \AHLTL are currently the only
logics for which practical model checking algorithms are
known~\cite{finkbeiner15algorithms,Hyperliveness,hsb21,bcbfs21}.
For \HyperLTL, the algorithms have been implemented in the model
checkers \MCHyper and bounded model checker \HyperQube~\cite{hbs21}.
\HyperLTL is limited to synchronous hyperproperties.
The \AHLTL model checking problem is known to be undecidable in general~\cite{bcbfs21}.
However, decidable fragments that can express observational determinism,
noninterference, and linearizability have been identified.
This paper generalizes \AHLTL by allowing nested trajectory quantifiers and due to the general undecidability result focuses on terminating programs.

FOL[E]~\cite{Hierarchy} can express a limited form of asynchronous hyperproperties.
As shown in~\cite{Hierarchy}, FOL[E] is subsumed by \HyperLTL with
additional quantification over predicates.
For $S1S[E]$ and $H_\mu$, the model checking problem is in general
undecidable; for $H_\mu$, two fragments, the $k$-synchronous,
$k$-context bounded fragments, have been identified for which model
checking remains decidable~\cite{DBLP:journals/pacmpl/GutsfeldMO21}.
Other logical extensions of \HyperLTL with asynchronous capabilities
are studied in~\cite{bozzelli21asynchronous}, including their
decidable fragments, but their model checking problems have not been
implemented and the relative expressive power with respect to other
asynchronous formalisms has not been studied.
%

\paragraph{Organization.} 
The rest of the paper is organized as follows. We generalize \AHLTL in 
Section~\ref{sec:ahltl}.
Section~\ref{sec:bmc} describes our bounded model checking algorithm while 
Section~\ref{sec:complexity} is dedicated to our complexity analysis.
Evaluation of our implementation results is presented in Section~\ref{sec:eval}. We conclude in 
Section~\ref{sec:concl}. Detailed proofs 
and descriptions of our case studies appear in the appendix.


	 \section{Extended Asynchronous HyperLTL}
\label{sec:ahltl}


\subsubsection{Preliminaries.}
%
Given a natural number $k  \in \naturals$, we use $[k]$ for the set 
$\{0,\ldots,k\}$.
Let $\AP$ be a set of {\em atomic propositions} and $\alphabet=2^\AP$
be the {\em alphabet}, where we call each element of $\alphabet$ a
{\em letter}.
A {\em trace} is an infinite sequence $\sigma=a_0a_1\cdots$ of letters from 
$\alphabet$.
We denote the set of all infinite traces by $\alphabet^\omega$.
We use $\sigma(i)$ for $a_i$ and $\sigma^i$ for the suffix
$a_ia_{i+1}\cdots$.
A {\em pointed trace} is a pair $(\sigma,p)$, where $p \in \naturals$ is a
natural number (called the {\em pointer}).
Pointed traces allow to traverse a trace by moving the pointer.
Given a pointed trace $(\sigma, p)$ and $n > 0$, we use
$(\sigma, p) + n$ to denote the resulting trace $(\sigma,p+n)$.
We denote the set of all pointed traces by
$\PTR = \{(\trace, p) \mid \trace \in \alphabet^\omega \, \text{ and } \, p
\in \naturals\}$.


A {\em Kripke structure} is a tuple $\krip = \ktuple$, where $\States$ is a set 
of states, $\state_{\init} \in \States$ is the initial state, $\trans \subseteq 
\States \times \States$ is a transition relation, and $L: S \rightarrow 
\alphabet$ is a labeling function on the states of $\krip$. We require that 
for each $\state \in \States$, there exists $\state' \in \States$, such 
that $(\state, \state') \in \trans$.\qed

A \emph{path} of a Kripke structure $\krip$ is an infinite sequence of states
$\state(0)\state(1)\cdots \in \States^\omega$, such that $\state(0) = 
\state_\init$ and $(\state(i), \state({i+1})) \in \trans$, for all $i \geq 0$.
A trace of $\krip$ is a sequence $\trace(0)\trace(1)\trace(2) \cdots 
\in \alphabet^\omega$, such that there exists a path $\state(0)\state(1)\cdots 
\in \States^\omega$ with $\trace(i) = L(\state(i))$ for all $i\geq 0$. 
We denote by $\Trace(\krip, \state)$ the set of all traces of $\krip$ with 
paths that start in state $\state \in \States$.

The directed graph $\kframe = \langle \States, \trans \rangle$ is 
called the {\em Kripke frame} of the Kripke structure $\krip$. A {\em loop} in 
$\kframe$ is a finite sequence $\state_0\state_1\cdots \state_n$, such that 
$(\state_i, \state_{i+1}) \in \trans$, for all $0 \leq i < n$, and $(\state_n, 
\state_0) \in \trans$. We call a Kripke frame {\em acyclic}, if the only loops 
are self-loops on terminal states, i.e., on states that have no other outgoing 
transition. Acyclic Kripke structures model terminating programs.

	 \subsubsection{Extended A-HLTL.}
The syntax of extended \AHLTL is:
\begin{alignat*}{4}
& \varphi ::= \exists \pi . \varphi && \mid \forall \pi. \varphi && \mid \Etau \tau. \varphi \mid \Atau \tau 
.\varphi \mid \psi \\
&\psi ::= \true && \mid a_{\pi, \tau} && \mid \lnot \psi \mid \psi_1 \Or \psi_2 
\mid \psi_1 \And \psi_2 \mid
\psi_1 \, \Until \, \psi_2 \mid \psi_1\release\psi_2 
\end{alignat*}
where $a \in \AP$, $\pi$ is a trace variable from an infinite supply
$\V$ of trace variables, $\tau$ is a {\em trajectory variable} from an
infinite supply $\J$ of trajectory variables (see formula 
$\varphi_\propNIndet$ in Section~\ref{sec:intro} for an example).
The intended meaning of $a_{\pi, \tau}$ is that proposition $a \in \AP$
holds in the current time in trace $\pi$ and {\em trajectory} $\tau$ (explained 
later).
Trace (respectively, trajectory) quantifiers $\exists\pi$ and $\forall\pi$ (respectively, $\Etau\tau$ and 
$\Atau\tau$) allow reasoning simultaneously about different traces (respectively, trajectories).
The intended meaning of $\Etau$ is that there is a trajectory that
gives an interpretation of the relative passage of time between the
traces for which the temporal formula that relates the traces is
satisfied.
Dually, $\Atau$ means that all trajectories satisfy the inner formula.
Given an \AHLTL formula $\varphi$, we use $\Paths(\varphi)$ (respectively, $\Trajs(\varphi)$) for 
the
set of trace (respectively, trajectory) variables quantified in $\varphi$.
A formula $\varphi$ is {\em well-formed} if for all atoms $a_{\pi, \tau}$ in
$\varphi$, $\pi$ and $\tau$ are quantified in $\varphi$ (\ie $\tau\in\Trajs(\varphi)$ and 
$\pi\in\Paths(\varphi)$) and no trajectory/trace variable is quantified twice in $\varphi$.
%
%
We use the usual syntactic sugar
$\false \defAs \neg \true$,
and
$\Event \varphi \defAs \true \, \Until \varphi$,
$\varphi_1 \into \varphi_2 \defAs \neg \varphi_1 \vee \varphi_2$, and
$\Always \varphi \defAs \neg \Event \neg \varphi$, etc.
We choose to add $\release$ (release) and $\And$ to the logic to enable
negation normal form (NNF).
As our BMC algorithm cannot handle formulas that are not invariant under 
stuttering, the {\em next} operator is not included.
%




\paragraph{\textit{Semantics.}}
%

%
A \emph{trajectory} $\traj: \traj(0)\traj(1)\traj(2)\cdots$ for a
formula $\varphi$ is an infinite sequence of subsets of $\Vars(\varphi)$, i.e., each $t_i \subseteq 
\Vars(\varphi)$, for all $i \ge 0$.
Essentially, in each step of the trajectory one or more of the traces
make progress or all may stutter.
A trajectory is {\em fair} for a trace variable $\pi \in \Vars(\varphi)$ if
there are infinitely many positions $j$ such that $\pi\in t(j)$.
A trajectory is fair if it is fair for all trace variables in
$\Vars(\varphi)$.
%
%
Given a trajectory $\traj$, by $\traj^i$, we mean the suffix
$\traj(i)\traj(i+1)\cdots$.
Furthermore, for a set of trace variables $\V$, we use $\tjall_\V$ for
the set of all fair trajectories for indices from $\V$.
%
%
%
%
We also use a {\em trajectory assignment}
$\tjass: \Trajs(\varphi) \partTo \tjall_{\Dom(\tjass)}$, where
$\Dom(\tjass)$ is the subset of $\Trajs(\varphi)$ for which $\tjass$
is defined.
Given a trajectory assignment $\tjass$, a trajectory variable $\tau$, and a trajectory
$t$, we denote by $\tjass[\tau \mapsto t]$ the assignment that coincides with $\tjass$ for every 
trajectory variable
except for $\tau$, which is mapped to $t$.

For the semantics of extended \AHLTL, we need asynchronous trace
assignments
$\Pi:\Paths(\varphi)\times\Trajs(\varphi)\Into T\times\Nat$ which map
each pair $(\pi,\tau)$ formed by a path variable and trajectory
variable into a pointed trace.
Given $(\Pi,\tjass)$ where $\Pi$ is an asynchronous trace assignment
and $\tjass$ a trajectory assignment, we use $(\Pi,\tjass)+1$ for the
successor of $(\Pi,\tjass)$ defined as $(\Pi',\tjass')$ where
$\tjass'(\tau)=\tjass(\tau)^1$, and $\Pi'(\pi,\tau)=\Pi(\pi,\tau)+1$
if $\pi\in \tjass(\tau)(0)$ and $\Pi'(\pi,\tau)=\Pi(\pi,\tau)$
otherwise.
Note that $\Pi$ can assign the same $\pi$ to different pointed traces
depending on the trajectory.
%
We use $(\trass,\tjass)+k$ as the $k$-th successor of $(\trass,\tjass)$.
Given an asynchronous trace assignment $\Pi$, a trace variable $\pi$, a trajectory variable $\tau$ a 
trace $\sigma$, and a pointer $p$, we denote by $\Pi[(\pi, \tau) \mapsto (\sigma,p)]$
the assignment that coincides with $\Pi$ for every pair  
except for $(\pi, \tau)$, which is mapped to $(\sigma,p)$.
The satisfaction of an \AHLTL formula $\varphi$ over a trace
assignment $\trass$, a trajectory assignment $\tjass$, and a set of
traces $T$ is defined as follows (we omit $\neg$, $\And$ and $\Or$
which are standard):
%
\[
  \begin{array}{rll@{\hspace{2.5em}}c@{\hspace{2.5em}}l}

 (\trass,\tjass)  &\models_T& \exists \trvar. \varphi & \text{iff} &   \text{for 
 some } \trace \in T: \\
 &&&& (\trass[(\pi,\tau) \mapsto (\trace, 0)], \tjass) \models_T \varphi \text{ for all $\tau$}\\
(\trass,\tjass)  &\models_T& \forall \pi. \varphi & \text{iff} & \text{for all } \trace  \in T:\\
&&&& (\trass [(\pi, \tau) \mapsto (\trace, 0)], \tjass) \models_T \varphi \text{ for all $\tau$}\\
(\trass,\tjass)  &\models_T& \Etau \tau. \psi & \text{iff} & \text{for some }\traj \in \tjall_{\Dom(\trass)}: 
(\trass,\tjass[\tau \mapsto t]) \models \psi\\
 (\trass,\tjass)  &\models_T& \Atau \tau.\psi & \text{iff} & \text{for all }\traj \in \tjall_{\Dom(\Pi)} 
 (\trass,\tjass[\tau \mapsto t]) \models \psi\\
(\trass,\tjass) &\models& a_{\trvar,\tjvar} & \text{iff} & a\in\sigma(n) \text{ where } 
(\sigma,n)=\Pi(\trvar,\tjvar)\\
%
%
%
%
%
(\trass,\tjass) &\models& \psi_1 \, \Until \psi_2 & \text{iff} &
                                                                 \text{for some } i\geq 0: (\trass,\tjass)+i\models \psi_2 \, \text{ and } \, \\
    &&&& \text{for all } j < i: (\trass,\tjass)+j \models \psi_1\\
    (\trass,\tjass) &\models& \psi_1 \, \release \psi_2 & \text{iff} & \text{for all } i\geq 0: (\trass,\tjass)+i\models \psi_2 \text{, or } \, \\
                  &&&&  \text{for some } i\geq 0: (\trass,\tjass)+i \models \psi_1 \text{ and } \\
                  &&&& \hspace{2em} \text{for all } j\leq{}i: (\trass,\tjass)+j \models \psi_2
\end{array}
\]
We say that a set $T$ of traces satisfies a sentence $\varphi$,
denoted by $T \models \varphi$, if $(\trass_\emptyset,\tjass_\emptyset) \models_T \varphi$.
We say that a Kripke structure $\krip$ satisfies an \AHLTL formula
$\varphi$ (and write $\krip \models \varphi$) if and only if we have
$\Trace(\krip, \States_\init)\models \varphi$.
An example is illustrated in Fig.~\ref{fig:example}.


\begin{figure}[t]
\centering
\scalebox{1.1}{	\scalebox{.55}{
\begin{tikzpicture}
	[-,>=stealth,shorten >=.5pt,auto, ellipse, minimum width=10pt, minimum height=10pt, node distance=.5cm, initial text={}]

	\node[initial, state] [font = \Large, text width=11mm, text centered, minimum  height=4em, minimum width= 
	11mm](0) at (0, 0) 
	{\small \code{h=0} \\  \code{l=0} \\ \vspace*{-1.6mm}\code{obs=`a'}};
	
	\node[state][font = \Large, text width=11mm, text centered, minimum height=5em,minimum width= 
	11mm] (1) at (2, 1.3)  
	{\small \code{h=0} \\ \code{l=1} \\ \vspace*{-1.6mm}\code{obs=`a'}};
	
	\node[state][font = \Large, text width=11mm, text centered, minimum height=5em,minimum width= 
	11mm] (2) at (4.3, 1.3) 
	{\small \code{h=0} \\ \code{l=0} \\ \vspace*{-1.6mm}\code{obs=`a'}};
	
	\node[state][font = \Large, text width=11mm, text centered, minimum height=5em,minimum width= 
	11mm] (3) at (2, -1.3) 
	{\small \code{h=1} \\ \code{l=1} \\ \vspace*{-1.6mm}\code{obs=`b'}};
	
	\node[state][font = \Large, text width=11mm, text centered, minimum height=5em,minimum width= 
	11mm] (4) at (4.3, -1.3) 
	{\small \code{h=0} \\ \code{l=0} \\ \vspace*{-1.6mm}\code{obs=`b'}};
	
	\node[state][font = \Large, text width=11mm, text centered, minimum height=5em,minimum width= 
	11mm] (5) at (6.3, 0) 
	{\small \code{h=0} \\ \code{l=0} \\ \vspace*{-1.6mm}\code{obs=`b'}};
	
	\draw[->]  
	(0) edge node (01 label) {} (1)
	(1) edge node (12 label) {} (2)
	(0) edge node (03 label) {} (3)
	(3) edge node (34 label) {} (4)
	(2) edge node (25 label) {} (5)
	(4) edge node (45 label) {} (5)
	(5) edge [loop right, looseness = 4] node (55 label) {} (5);
	
\end{tikzpicture}
}
\hspace*{-7mm}
\newcommand{\myfontsize}{\small}
\scalebox{.55}{
\begin{tikzpicture}
	\tikzstyle{eclp}=[draw=black, ellipse, minimum width=50pt, minimum height=50pt,
	align=center]
	\begin{scope}
		\node [eclp](a1) 
		{\myfontsize \code{h=0} \\ \myfontsize \code{l=0} \\ \myfontsize \code{obs=`a'}};		
		\node [left=30pt] at (a1) {\code{$t_1$}};
		
		\node[eclp, left=150pt, right=40pt] (a2) at (a1) 
		{\myfontsize \code{h=0} \\ \myfontsize \code{l=1} \\ \myfontsize \code{obs=`a'}};
		
		\node[eclp, left=150pt, right=40pt] (a3) at (a2) 
		{\myfontsize \code{h=0} \\ \myfontsize \code{l=0} \\ \myfontsize \code{obs=`a'}};		
		
		\node[eclp, left=150pt, right=40pt] (a4) at (a3) 
		{\myfontsize \code{h=0} \\ \myfontsize \code{l=0} \\ \myfontsize \code{obs=`b'}};		

		\draw[->] (a1) -- (a2);
		\draw[->] (a2) -- (a3);
		\draw[->] (a3) -- (a4);

	\end{scope}	
	
	\begin{scope}[yshift=-2.8cm]
		\node [eclp](b1) 
		{\myfontsize \code{h=0} \\ \myfontsize \code{l=0} \\ \myfontsize \code{obs=`a'}};		
		\node [left=30pt] at (b1) {\code{$t_2$}};
		
		\node[eclp, left=150pt, right=40pt] (b2) at (b1) 
		{\myfontsize \code{h=1} \\ \myfontsize \code{l=1} \\ \myfontsize \code{obs=`b'}};		
		
		\node[eclp, left=150pt, right=40pt] (b3) at (b2) 
		{\myfontsize \code{h=0} \\ \myfontsize \code{l=0} \\ \myfontsize \code{obs=`b'}};		
		
		\node[eclp, left=150pt, right=40pt] (b4) at (b3) 
		{\myfontsize \code{h=0} \\ \myfontsize \code{l=0} \\ \myfontsize \code{obs=`b'}};		
		
		\draw[->] (b1) -- (b2);
		\draw[->] (b2) -- (b3);
		\draw[->] (b3) -- (b4);
	\end{scope}
	
	\begin{scope}[transform canvas={xshift=-.7em}]
		\draw[-, dashed, color=red] (a1) -- (b1);
		\draw[-, dashed, color=red] (a2) -- (b2);
		\draw[-, dashed, color=red] (a3) -- (b3);
		\draw[-, dashed, color=red] (a4) -- (b4);
	\end{scope}

	\begin{scope}[transform canvas={xshift=.1em}]
		\draw[-, dashed, color=blue] (a1) -- (b1);
		\draw[-, dashed, color=blue] (a2) -- (b1);
		\draw[-, dashed, color=blue] (a3) -- (b1);
		\draw[-, dashed, color=blue] (a4) -- (b2);
		\draw[-, dashed, color=blue] (a4) -- (b3);
		\draw[-, dashed, color=blue] (a4) -- (b4);
	\end{scope}

\end{tikzpicture}
} }
\caption{Kripke structure $\krip$ (left) and the two traces $t_1$ and 
$t_2$ of $\krip$ (right), $\krip \models \varphi_\propNIndet$ 
	but $\krip \not\models \varphi_\propNI$.
}
\label{fig:example}
\end{figure}



\section{Bounded Model Checking for A-HLTL}
\label{sec:bmc}

\newcolumntype{C}{>{$}c<{$}}
\newcolumntype{L}{>{$}l<{$}}
\newcolumntype{R}{>{$}r<{$}}
\newcolumntype{F}{>{$}X<{$}}

We first introduce the bounded semantics of \AHLTL (for at most one 
trajectory quantifier alternation but arbitrary trace quantifiers) which will be 
used to generate queries to a QBF solver to aid solving the BMC problem.
The main result of this section is Theorem~\ref{thm:qbf-mc-unroll-new} 
which provides decision procedures for model checking \AHLTL for 
terminating systems.
%

\subsection{Bounded Semantics of A-HLTL}
The bounded semantics corresponds to the exploration of the system up
to a certain bound.
In our case, we will consider two bounds $k$ and $m$ (with $k\leq
m$).
The bound $k$ corresponds to the {\em maximum depth} of the unrolling of the
Kripke structures and $m$ is the {\em bound on trajectories length}.
We start by introducing some auxiliary functions and predicates, for a
given trace assignment and $(\Pi,\Gamma)$.
First, the family of functions
$\Pos_{\pi,\tau}:\{0\ldots{}m\}\Into\Nat$.
The meaning of $\Pos_{\pi,\tau}(i)$ provides how many times $\pi$
has been selected in $\{\tau(0),\ldots,\tau(i)\}$.
We assume that Kripke structures are equipped with an atomic
proposition $\Halt$ (one per trace variable $\pi$) which encodes
whether the state is a halting state.
Given $(\Pi,\Gamma)$ we consider the predicate $\Halted$ that holds
whenever for all $\pi$ and $\tau$, $\Halt\in\sigma(j)$ for
$(\sigma,j)=\Pi(\pi,\tau)$.
In this case we write $(\Pi,\Gamma,n)\models\Halted$.

We define two bounded semantics which only differ in how they inspect
beyond the $(k,m)$ bounds: $\HPmodels_{k,m}$, called the \emph{halting
  pessimistic semantics} and $\HOmodels_{k,m}$, called the
\emph{halting optimistic semantics}.
%
%
We start by defining the bounded semantics of the quantifiers.
\[
\begin{array}{lll@{\hspace{1.8em}}c@{\hspace{1.8em}}lr}
(\Pi,\Gamma,0) & \models_{k,m} & \exists \pi.\ \psi & \hspace{0cm}\text{ iff 
} 
\hspace{1cm} & \text{there is a } \sigma \in T_\pi \text{, such that  for all } \tau\\
&&&&(\Pi[(\pi,\tau)\Into(\sigma,0)],\Gamma,0) \models_{k,m} \psi & 
\hspace{1cm} (1)\\
(\Pi,\Gamma,0) &\models_{k,m} & \forall \pi.\ \psi & \text{ iff } & \text{for all } 
\sigma \in T_\pi, \text{ for all } \tau:\\
&&&& (\Pi[(\pi,\tau)\Into(\sigma,0)],\Gamma,0)  \models_{k,m} \psi & (2)\\
(\Pi,\Gamma,0) & \models_{k,m} & \Etau \tau.\ \psi & \text{ iff } & 
\text{there is a } t \in \tjall_{\Dom(\trass)}:\\
&&&& (\Pi,\Gamma[\tau\Into{}t],0) \models_{k,m} \psi & (3)\\
(\Pi,\Gamma,0) & \models_{k,m} & \Atau \tau.\ \psi & \text{ iff } & \text{for 
all } 
t \in  \tjall_{\Dom(\trass)}:&\\
&&&& (\Pi,\Gamma[\tau\Into{}t],0) \models_{k,m} \psi & (4)

\end{array}
\]
For the Boolean operators, for $i\leq{} m$:

\noindent\begin{tabularx}{\columnwidth}{L@{\hspace{2.5em}}C@{\hspace{2.5em}}FR}\\[-0.3em]
(\Pi, \Gamma,i) \models_{k,m} \tru & &  & (5) \\
(\Pi, \Gamma,i) \models_{k,m}  a_{\pi,\tau} & \text{iff} & a \in (\sigma,j) \text{ 
where } & \\ && \hspace{0.2em} (\sigma,j)=\Pi(\pi,\tau)(i) \text{ and } 
j\leq{}k& (6) \\
(\Pi, \Gamma,i) \models_{k,m}  \neg a_{\pi,\tau} & \text{iff} & a \not\in (\sigma,j) \text{ where } & \\ && \hspace{0.2em}(\sigma,j)=\Pi(\pi,\tau)(i) \text{ and } j\leq{}k & (7)\\
%
%
(\Pi, \Gamma,i) \models_{k,m}  \psi_1 \vee \psi_2 & \text{iff} & (\Pi,\Gamma,i) \models_{k,m}
\psi_1 \text{ or } (\Pi,\Gamma,i) \models_{k,m} \psi_2 & (8) \\
(\Pi, \Gamma,i) \models_{k,m}  \psi_1 \wedge \psi_2 & \text{iff} & (\Pi,\Gamma,i) \models_{k,m} 
\psi_1 \text{ and } (\Pi,\Gamma,i) \models_{k,m} \psi_2 & (9)\\[0.6em]
\end{tabularx}

For the temporal operators, we must consider the cases of falling of
the paths (beyond $k$) and falling of the traces (beyond $m$).
We define the predicate $\off$ which holds for $(\Pi,\Gamma,i)$ if for
some $(\pi,\tau)$, $\Pos_{\pi,\tau}(i)>k$ and
$\Halt_\pi\notin\sigma(k)$ where $\sigma$ is the trace assigned to $\pi$.
%
Note that $\Halted$ implies that $\off$ does not hold because all
paths (including those at $k$ or beyond) satisfy $\Halt$.
%

We define two semantics that differ on how to interpret when the end
of the unfolding of the traces and trajectories is reached.
The {\em halting pessimistic} semantics, denoted by $\HPmodels_{k,m}$ take
$(1)$-$(9)$ above 
and add $(10)$-$(13)$ together with $(\Pi,\Gamma,i)\not\models_{k,m}\off$. 
%
Rules $(10)$ and $(11)$ define the semantics of the temporal operators
for the case $i<m$, that is, before the end of the unrolling of the
trajectories (recall that we do not consider $\Next$):

\noindent\begin{tabularx}{\columnwidth}{R@{\hspace{1em}}L@{\hspace{1em}}C@{\hspace{0.5em}}FR}\\[-0.3em]
	(\Pi,\Gamma,i) \models_{k,m} & \psi_1 \Until \psi_2 & \text{iff} &  (\Pi,\Gamma,i) 
	\models_{k,m}\psi_2\text{, or } (\Pi,\Gamma,i) \models_{k,m}\psi_1 \text{, 
	and} & \\ &&&  (\Pi,\Gamma,i)+1 \models_{k,m}\psi_1 \Until \psi_2  & (10) 
\end{tabularx}

\noindent\begin{tabularx}{\columnwidth}{R@{\hspace{1em}}L@{\hspace{1em}}C@{\hspace{0.5em}}FR}\\[-0.3em]
  (\Pi,\Gamma,i) \models_{k,m} & \psi_1 \release \psi_2 & \text{iff} & (\Pi,\Gamma,i) 
  \models_{k,m}\psi_2 \text{, and } (\Pi,\Gamma,i) \models_{k,m}\psi_1 \text{, or}& \\
  &&&  (\Pi,\Gamma,i)+1 \models_{k,m}\psi_1\release \psi_2  & (11)\\[0.6em]
  \end{tabularx}
  For the case of $i=m$, that is, at the bound of the
  trajectory:
  
  \noindent\begin{tabularx}{\columnwidth}{R@{\hspace{1em}}L@{\hspace{1em}}C@{\hspace{0.6em}}FR}\\[-0.3em]
    (\Pi,\Gamma,m) \HPmodels_{k,m} & \psi_1 \Until \psi_2 & \text{iff} & (\Pi,\Gamma,m)\models_{k,m} \psi_2 & (12) \\
    (\Pi,\Gamma,m) \HPmodels_{k,m} & \psi_1 \release \psi_2 & \text{iff} & (\Pi,\Gamma,m) \models_{k,m}\psi_1\And\psi_2 \text{, or}& \\
    &&& (\Pi,\Gamma,m) \models_{k,m}\Halted\And\psi_2 & (13) \\
  \end{tabularx}

  The {\em halting optimistic} semantics, denoted by $\HOmodels_{k,m}$ take
  rules $(1)$-$(11)$ and $(12')$-$(13')$, but now if
  $(\Pi,\Gamma,i)\HOmodels_{k,m}\off$ then
  $(\Pi,\Gamma,i)\HOmodels_{k,m}\varphi$ holds for every formula.
  Again, rules $(10)$ and $(11)$ define the semantics of the temporal
  operators for the case $i<m$.
Then, for $i=m$:

  \noindent\begin{tabularx}{\columnwidth}{R@{\hspace{0.5em}}L@{\hspace{0.6em}}C@{\hspace{0.6em}}FR}\\[-0.3em]
    (\Pi,\Gamma,m) \HOmodels_{k,m} & \psi_1 \Until \psi_2 & \text{iff} & 
    (\Pi,\Gamma,m) \models_{k,m}\psi_2 \text{, or}& \\
    &&& (\Pi,\Gamma,m) \not\models_{k,m}\Halted\And\psi_1 & (12')\\
    (\Pi,\Gamma,m) \HOmodels_{k,m} & \psi_1 \release \psi_2 & \text{iff} &    
    (\Pi,\Gamma,m) \models_{k,m}\psi_2 & (13') \\[0.6em]
  \end{tabularx}
%
	
  As the semantics introduced in~\cite{hsb21} for the case of
  \HyperLTL, the pessimistic semantics capture the case where we
  assume that pending eventualities will not become true in the future
  after the end of the trace (this is also assumed in \LTL BMC).
  Dually, the optimistic semantics assume that all pending
  eventualities at the end of the trace will be fulfilled.
  Therefore, the following hold.

\begin{lemma}
  \label{lem:incr}
  Let $k\leq k'$ and $m\leq m'$.
  \begin{compactenum}
  \item If $(\Pi,\Gamma,0)\HPmodels_{k,m}\varphi$, then
    $(\Pi,\Gamma,0)\HPmodels_{k',m'}\varphi$.
  \item If $(\Pi,\Gamma,0)\not\HOmodels_{k,m}\varphi$, then
    $(\Pi,\Gamma,0)\not\HOmodels_{k',m'}\varphi$.
    \end{compactenum}
\end{lemma}

\begin{lemma}
  \label{lem:infinite-inference}
  The following hold for every $k$ and $m$,
    \begin{compactenum}
  \item If $(\Pi,\Gamma,0)\HPmodels_{k,m}\varphi$, then
    $(\Pi,\Gamma,0)\models\varphi$.
  \item If $(\Pi,\Gamma,0)\not\HOmodels_{k,m}\varphi$, then
    $(\Pi,\Gamma,0)\not\models\varphi$.
  \end{compactenum}    
\end{lemma}

\subsection{From Bounded Semantics to QBF Solving}
%

Let $\krip$ be a Kripke structure and $\varphi$ be an \AHLTL formula.
Based on the bounded semantics introduced previously, our main
approach is to generate a QBF query (with bounds $k$, $m$), which
can use either the pessimistic or the optimistic semantics.
We use $\QBF{\krip,\varphi}^\HPES_{k,m}$ if the pessimistic semantics
are used and $\QBF{\krip,\varphi}^\HOPT_{k,m}$ if the optimistic
semantics are used.
Our translations will satisfy that
\begin{enumerate}[(1)]
\item if $\QBF{\krip,\varphi}^\HPES_{k,m}$ is SAT, then $\krip\models\varphi$;
\item if $\QBF{\krip,\varphi}^\HOPT_{k,m}$ is UNSAT, then $\krip\not\models\varphi$;
\item if the Kripke structure is unrolled to the diameter and the
  trajectories up to a maximum length (see below), then
  $\QBF{\krip,\varphi}^\HPES_{k,m}$ is SAT if and only if
  $\QBF{\krip,\varphi}^\HOPT_{k,m}$ is SAT.
\end{enumerate}
The first step to define $\QBF{\krip,\varphi}^\HOPT_{k,m}$ and
$\QBF{\krip,\varphi}^\HPES_{k,m}$ is to encode the unrolling of the models
up-to a given depth $k$.
For a path variable $\pi$ corresponding to Kripke structure
$\krip$, we introduce $(k+1)$ copies ($x^0,\ldots,x^k$) of the
Boolean variables that define the state of $\krip$ and use the
initial condition $I$ and the transition relation $R$ of $\krip$
to relate these variables.
For example, for $k=3$, we unroll the transition relation up-to
$3$ as follows:
\vspace*{-.1em}
$$
\sem{\krip}_3 = I(x^0) \land R(x^0, x^1) \land R(x^1, x^2) \land R(x^2, x^3).
$$
\vspace*{-2.05em}

\begin{wrapfigure}{r}{.425 \textwidth}
	\newcommand{\spacing}{\hspace*{1mm}}
	\vspace*{-10mm}
		\input{figs/encoding}
		\vspace*{-4mm}	
		\caption{Variables for encodings of the {blue trajectory} in Fig.~\ref{fig:example}, where 
			{green variables} are $\true$ and {gray variables} are $\false$.
		}
		\label{fig:encodings}
		\vspace*{-7mm}	
	\end{wrapfigure}
\paragraph{Encoding positions.}
\label{para:encoding}
For each trajectory variable $\tjvar$ and given the bound $m$ on the
unrolling of trajectories, we add $\Paths(\varphi)\times(m+1)$ variables $t_{\pi}^0\ldots 
t_{\pi}^m$, for 
each $\pi$.
The intended meaning of $t_{\pi}^j$ is that $t_{\pi}^j$ is true
whenever $\pi\in{}t(j)$, that is, when $t$ dictates that $\pi$ moves
at time instant $j$.
In order to encode sanity conditions on trajectories, that are crucial
for completeness, it is necessary to introduce a family of variables
that captures how much $\pi$ has moved according to $\tau$ after $j$
steps.
There is a variable $\Pos$ for each trace variable $\pi$, each
trajectory $\tau$ and each $i\leq{}k$ and $j\leq{}m$.
We represent this variable by $\Pos^{i,j}_{\pi,\tau}$.
The intention is that $\Pos$ is true whenever after $j$ steps
trajectory $\tau$ has dictated that trace $\pi$ progresses precisely
$i$ times. 
Fig.~\ref{fig:encodings} shows encodings $t_{\pi}^j$ and $\Pos^{i,j}_{\pi,\tau}$ for the traces 
 w.r.t. the blue trajectory, $\tau'$ in Fig.~\ref{fig:example}.
We will use the auxiliary definitions (for $i\in\{0\ldots{}k\}$ and
$j\in\{0\ldots{}m\}$) to force that the path $\pi$ has moved to position $i$
after $j$ moves from the trajectory and that $\pi$ has not fallen off
the trace (and does not change position when the paths fall off the trace):
\begin{align*}
  \SETPOS_{\pi,\tau}^{i,j} & \DefinedAs \Pos_{\pi,\tau}^{i,j} \And \bigwedge_{n \in \{0..k\} \setminus \{i\}}\neg\Pos_{\pi,\tau}^{n,j}\And\neg\off^j_{\pi,\tau} \\
  \NOPOS_{\pi,\tau}^{j} & \DefinedAs \off^j_{\pi,\tau}\And \bigwedge_{n\in\{0..k\}}\neg\Pos_{\pi,\tau}^{n,j} 
\end{align*}
Initially,
$
  I_\Pos\DefinedAs\bigwedge_{\pi,\tau}\SETPOS^{0,0}_{\pi,\tau}
$,
where $\pi \in \Traces(\varphi)$ and $\tau \in \tjall_{\Dom(\trass)}$.
$I_\Pos$ captures that all paths are initially at position $0$.
Then, for every step $j\in\{0\ldots m\}$, the following formulas relate
the values of $\Pos$ and $\off$, depending on whether trajectory
$\tau$ moves path $\pi$ or not (and on whether $\pi$ has reached the
end $k$ or halted):
\begin{align*}
 \STEP^j_{\pi,\tau} & \DefinedAs \bigwedge_{i\in\{0..k-1\}}\big(\Pos^{i,j}_{\pi,\tau} \And t_\pi^{j} \Into \SETPOS^{i+1,j+1}_{\pi,\tau}\big)\\
\end{align*}   
\begin{align*}
 \STUTTERS^j_{\pi,\tau}& \DefinedAs \bigwedge_{i\in\{0..k\}}\big(
   \Pos^{i,j}_{\pi,\tau} \And \neg t_\pi^{j} \Into \SETPOS^{i,j+1}_{\pi,\tau}\big)  \\ 
 \ENDS^{j}_{\pi,\tau} &\DefinedAs (\Pos^{k,j}_{\pi,\tau} \And t_\pi^j) \Into \big((\neg{\Halt\,}^k_{\pi} 
 \Into \NOPOS^{j+1}_{\pi,\tau}) \And ({\Halt\,}^k_{\pi} \Into \SETPOS^{k,j+1}_{\pi,\tau})\big)
\end{align*}
Then the following formula captures the correct assignment to the the
$\Pos$ variables, including the initial assignment:
\[
  \PPos\DefinedAs I_\Pos \And \bigwedge_{j\in\{0..m\}}\bigwedge_{\pi,\tau} (\STEP_{\pi,\tau}^j \And \STUTTERS_{\pi,\tau}^j\And \ENDS_{\pi,\tau}^j)
\]
	For example, Fig.~\ref{fig:encodings} (w.r.t. Fig.~\ref{fig:example})
	encodes the  blue trajectory $(\tau')$ of $\pi$ (i.e., $t_1$) and
	$\pi'$ (i.e., $t_2$)  as follows.  
	First, 
	for $j\in[0,3)$, 
	it advances $t_1$ and
	stutters $t_2$. 
	%
	Therefore, 
	${t_\pi^{0}} , {t_\pi^{1}}, { t_\pi^{2}}$ are $\true$ and 
	${t_{\pi'}^{0}}, {t_{\pi'}^{1}}, { t_{\pi'}^{2}}$ are $\false$.
	Notice that for $\Pos$ encodings, the $\pi$ position advances according to $\STEP^j_{\pi,\tau'}$
	(i.e., 
	${\Pos_{\pi,\tau'}^{0,0}}, 
	{\Pos_{\pi,\tau'}^{1,1}}, 
	{\Pos_{\pi,\tau'}^{2,2}}, 
	{\Pos_{\pi,\tau'}^{3,3}}$);
	while $\pi'$ stutters according to $ \STUTTERS^j_{\pi',\tau'}$ 
	(i.e., 
	$ {\Pos_{\pi',\tau'}^{0,0}}, 
	{\Pos_{\pi',\tau'}^{0,1}}, 
	{\Pos_{\pi',\tau'}^{0,2}}, 
	{\Pos_{\pi',\tau'}^{0,3}}$).
	Then, for $j \in [3,5]$, it alternatively advances $t_2$ which makes 
	${ t_\pi^{3}}, {t_\pi^{4}},  {t_\pi^{5}}$ $\false$ and 
	$ { t_{\pi'}^{3}}, { t_{\pi'}^{4}}, { t_{\pi'}^{5}}$ $\true$.
	Similarly, the movements becomes 
	${\Pos_{\pi,\tau'}^{3,4}}, 
	{\Pos_{\pi,\tau'}^{3,5}}, 
	{\Pos_{\pi,\tau'}^{3,6}}
	$ 
	and 
	${\Pos_{\pi',\tau'}^{1,4}}, 
	{\Pos_{\pi',\tau'}^{2,5}}, 
	{\Pos_{\pi',\tau'}^{3,6}}$.
	At the halting point (i.e., $j = k$), both trajectory trigger $\ENDS^{j}$ and do not advance anymore.

\paragraph{Encoding the inner LTL formula.}
We will use the following auxiliary predicates:
\[
  {\Halted\,}^j \DefinedAs \bigwedge_{\tau} {\Halted\,}^j_\tau
  \hspace{5em} \off^j\DefinedAs\bigvee_{\pi,\tau}\off^j_{\pi,\tau}
\]
We now give the encoding for the inner temporal formulas for a fix
unrolling $k$ and $m$ as follows.
For the atomic and Boolean formulas, the following translations are
performed for $j\in\{0\ldots{}m\}$.

\begin{tabularx}{\columnwidth}{LCFR}\\
  \QBF{p_{\pi,\tau}}_{k,m}^j & := & \bigvee_{i\in\{0..k\}}(\Pos^{i,j}_{\pi,\tau} \And p^i_\pi) & (14) \\
  \QBF{\neg{}p_{\pi,\tau}}_{k,m}^j & := & \bigvee_{i\in\{0..k\}}(\Pos^{i,j}_{\pi,\tau} \And \neg{}p^i_\pi) & (15)\\
  \QBF{\psi_1\Or\psi_2}_{k,m}^j & := &     \QBF{\psi_1}_{k,m}^j\Or    \QBF{\psi_2}_{k,m}^j & (16)\\
  \QBF{\psi_1\And\psi_2}_{k,m}^j & := &    \QBF{\psi_1}_{k,m}^j\And    \QBF{\psi_2}_{k,m}^j &(17) \\
\end{tabularx}
\vspace{2mm}

The halting pessimistic semantics translation uses $\QBF{\cdot}_\HPES$,
taking $(14)$-$(17)$ and $(18)$-$(21)$ below.
For the temporal operators and $j<m$:
%

\begin{tabularx}{1.00\columnwidth}{L@{}C@{}FR}\\
  \QBF{\psi_1\Until\psi_2}_{k,m}^j & :=& \neg\off^j \And \big(\QBF{\psi_2}_{k,m}^j \Or ( \QBF{\psi_1}_{k,m}^j\And\QBF{\psi_1\Until\psi_2}_{k,m}^{j+1})\big) & (18)\\
\QBF{\psi_1\release\psi_2}_{k,m}^j &:=& 
\neg\off^j \And 
\big( \QBF{\psi_2}_{k,m}^j \And ( \QBF{\psi_1}_{k,m}^j\Or\QBF{\psi_1\release\psi_2}_{k,m}^{j+1})\big) & (19)  \\[0.6em]
\end{tabularx}
 
\noindent For $j=m$:

\begin{tabularx}{1.00\columnwidth}{L@{}C@{}FR}\\
  \QBF{\psi_1\Until\psi_2}_{k,m}^m & :=& \QBF{\psi_2}_{k,m}^m& (20)\\
  \QBF{\psi_1\release\psi_2}_{k,m}^m & :=& \big(\QBF{\psi_1}_{k,m}^m \And \QBF{\psi_2}_{k,m}^m\big)\Or\big({\Halted\,}^m\And \QBF{\psi_2}_{k,m}^m\big)& (21)\\[0.6em]
\end{tabularx}

The halting optimistic semantics translation uses $\QBF{\cdot}_\HOPT$,
taking $(14)$-$(17)$ and $(18')$-$(21')$ as follows,
For the temporal operators and $j<m$:
%

\begin{tabularx}{1.00\columnwidth}{L@{}C@{}FR}\\
  \QBF{\psi_1\Until\psi_2}_{k,m}^j & :=& \off^j \Or \big(\QBF{\psi_2}_{k,m}^j \Or ( \QBF{\psi_1}_{k,m}^j\And\QBF{\psi_1\Until\psi_2}_{k,m}^{j+1})\big) & (18')\\
  \QBF{\psi_1\release\psi_2}_{k,m}^j &:=& \off^j \Or 
 	\big( \QBF{\psi_2}_{k,m}^j \And ( \QBF{\psi_1}_{k,m}^j\Or\QBF{\psi_1\release\psi_2}_{k,m}^{j+1})\big) & (19') \\[0.6em]
\end{tabularx}
 
\noindent For $j=m$:
  
\begin{tabularx}{1.00\columnwidth}{L@{}C@{}FR}\\
  \QBF{\psi_1\Until\psi_2}_{k,m}^m & :=& \QBF{\psi_2}_{k,m}^m \Or\big({\Halted\,}^m\And \QBF{\psi_1}_{k,m}^m\big) & (20')\\
  \QBF{\psi_1\release\psi_2}_{k,m}^m & :=& \QBF{\psi_2}_{k,m}^m& (21')\\[0.6em]
\end{tabularx}

\paragraph{Combining the encodings.}
Let $\varphi$ be a \AHLTL formula of the form \linebreak
$\varphi =
\quant_A\pi_A.\dots.\quant_Z\pi_Z.\quant_a\tau_a.\dots.\quant_z\tau_z.\psi$.
%
Combining all the components, the encoding of the \AHLTL BMC problem
into QBF, for bounds $k$ and $m$ is:
\begin{align*}
	 \QBF{\krip, \varphi}_{k,m} = \quant_A\overline{x_A}.\cdots.\quant_Z\overline{x_Z}.
	\quant_a\overline{t_a}.\cdots.\quant_z\overline{t_z}.\
	\exists \overline{\Pos}.\ \exists\overline{\off}.   \\
	 \Big( \QBF{\krip}_k
	\circ_A \cdots \QBF{\krip}_k \circ_Z
	( \PPos \And \Enc(\psi) )
	\Big)
\end{align*}
where $\circ_A=\Into$ if $\quant_A=\forall$ (and $\circ_A=\And$ if
$\quant_A=\exists$), and $\circ_B$, $\ldots$ are defined similarly.
The sets $\overline{\Pos}$ is the set of variables
$\Pos^{i,j}_{\pi,\tau}$ that encode the positions and
$\overline{\off}$ is the set of variables $\off^j_{\pi,\tau}$ that
encode when a trace progress has fallen off its unrolling limit.
We next define the encoding $\Enc(\psi)$ of the temporal formula $\psi$.

\paragraph{Encoding formulas with up to 1 trajectory quantifier
  alternations}
We consider the encoding into QBF of formulas with zero and one
quantifier alternation separately.
In the following, we say that at position $j$ a collection of
trajectories $U$ ``moves'' whenever either all trajectories have moved
all their paths to the halting state, or at least one of the
trajectories in $U$ makes one of the non-halted path move at position
$j$.
Formally,
\[
  \MOVES^j_U\DefinedAs {\Halted\,}^j_U \Or \bigvee_{\tau\in{}U,\pi} (t^j_\pi \And \neg{\Halt\,}_{\pi, \tau}^{j}) 
\]
\begin{itemize}
\item $\Etau^+ U. \psi$: In this case, the formula generated for $ \Enc(\psi)$ is 
  \[ (\bigwedge_{j\in\{0\ldots{}m\}}\MOVES^j_U) \And \QBF{\psi}^0_{k,m} \]
  This is correct since the positions at which all
  trajectories stutter all paths can be removed (obtaining a
  satisfying path), we can restrict the search to non-stuttering
  trajectory steps.
\item $\Atau^+ U. \psi$: In this case, the formula generated for $ \Enc(\psi)$ is
  \[  (\bigwedge_{j\in\{0\ldots{}m\}}\MOVES^j_U) \Into \QBF{\psi}^0_{k,m} \]
  The reasoning is similar as the previous case.
\item $\Atau^+ U_A\Etau^+ U_E. \psi$: In this case, the formula
  generated for $ \Enc(\psi)$ is
  \[ (\bigwedge_{j \in\{0\ldots{}m\}}\MOVES^j_{U_A}) \Into \big(
    \bigwedge_{j \in\{0\ldots{}m\}}({\Halted\,}^j_{U_A} \Into 
    \MOVES^j_{U_E})\And\QBF{\psi}^0_{k,m}\big) \]
  Universally quantified trajectories must explore all trajectories,
  which must be responded by the existential trajectories.
  Assume there is a strategy for $U_E$ for the case that universal
  trajectories $U_A$ never stutter at any position.
  This can be extended into a strategy for the case where $U_A$ can
  possible stutter, by adding a stuttering step to the
  $U_E$ trajectories at the same position.
  This guarantees the same evaluation.
  Therefore, we restrict our search for the outer $U_A$ to
  non-stuttering trajectories.
  Finally, $U_E$ is obliged to move after $U_A$ has halted all paths
  to prevent global stuttering.
\item $\Etau^+ U_E\Atau^+ U_A. \psi$: In this case, the
  formula generated for $ \Enc(\psi)$ is similar,
  \[ \big(\bigwedge_{j \in\{0\ldots{}m\}}\MOVES^j_{U_E}\big) \land  
  \big(\bigwedge_{j\in\{0\ldots{}m\}}({\Halted\,}^j_{U_E} \Into 
  \MOVES^j_{U_A})\Into\QBF{\psi}^0_{k,m}\big) \]
  The rationale for this encoding is the following.
  It is not necessary to explore a non-moving step $j$ for the
  existentially quantified trajectories $U_E$ because if this
  stuttering step is successful it must work for all possible moves of
  the $U_A$ trajectories at the same time step $j$.
  This includes the case that all trajectories in $U_A$ make all paths
  stutter (which, if we remove $j$ one still has all the legal
  trajectories for $U_A$).
  Since the logic does not contain the next operator, the evaluation
  for the given $U_E$ and one of the trajectories for $U_A$ that
  stutter at $j$ will be the same as for $j+1$ for all logical formulas.
  Therefore, the trajectory that is obtained from removing step $j$
  from $U_E$ is still a satisfying trajectory assignment.
  It follows that if there is a model for $U_E$ there is a model that
  does not stutter.
  Finally, after all paths have halted according to the $U_E$
  trajectories, a step of $U_A$ that stutters all paths that have not
  halted can be removed because, again the evaluation is the same in
  the previous and subsequent state.
  It follows that if the formula has a model, then it has a model
  satisfying the encoding.
\end{itemize}

\begin{theorem}
  \label{thm:qbf-mc-unroll-new}
  Let $\varphi$ be an \AHLTL formula with at most one trajectory
  quantifier alternation, let $K$ be the maximum depth of a Kripke
  structure and let $M=K \times |\Paths(\varphi)| \times |\Trajs(\varphi)|$.
  Then, the following hold:
  \begin{compactitem}
  \item $\QBF{\krip,\varphi}_{K,M}^\HPES$ is satisfiable if and only if
    $\krip\models\varphi$.
  \item $\QBF{\krip,\varphi}_{K,M}^\HOPT$ is satisfiable if and only if
        $\krip\models\varphi$.
  \end{compactitem}    
\end{theorem}
Theorem~\ref{thm:qbf-mc-unroll-new} provides a model checking decision
procedure. An alternative decision procedure is to iteratively increase the bound
of the unrollings and invoke both semantics in parallel until the
outcome coincides.


 	 \section{Complexity of A-HLTL Model Checking for Acyclic Frames}
\label{sec:complexity}
Our goal in this section is to analyze the complexity of the \AHLTL model checking problem in the 
size of an acyclic Kripke structure.
\vspace*{-5mm}
\subsubsection{Problem Formulation.}
We use \MC{\sf Fragment}~to distinguish different variations of the problem, where
\comp{MC} is the model checking decision problem, i.e., whether or not $\krip \models 
	\varphi$, 
	and \comp{Fragment} is one of the following for $\varphi$:
	
	\begin{compactitem}
		\item `$[\exists(\exists/\forall)^+\Atau/\Etau]^k$', for $k\geq 0$, denotes the fragment with a 
		lead existential trace quantifier, one outermost universal or existential trajectory quantifier, and 
		$k$ quantifier alternations (counting {\em all} quantifiers), where $k=0$ means the existential
		alternation-free fragment `$\exists^+\Etau^+$'. Fragment 
		`$[\forall(\forall/\exists)^+\Atau/\Etau]^k$' is defined similarly, where $k=0$ is the universal 
		alternation-free fragment `$\forall^+\Atau^+$'. 
		
		\item  Fragments 
		`$[\exists(\exists/\forall)^+(\Etau^+\Atau^+/\Atau^+\Etau^+/\Etau\Etau^+/\Atau\Atau^+)]^k$',
		for $k \geq 1$
		denotes the fragment with a lead existential trace quantifier, multiple outermost trajectory 
		quantifiers with at most one alternation, and $k$ quantifier alternations (counting {\em 
		all} quantifiers), where $k=1$ means fragment `$\exists\Etau\Atau$'. 
		Fragment 		
		`$[\forall(\forall/\exists)^+(\Etau^+\Atau^+/\Atau^+\Etau^+/\Etau\Etau^+/\Atau\Atau^+)]^k$'
		 is 
		defined similarly, where $k=1$ means fragment `$\forall\Atau\Etau$'.
		
%
	\end{compactitem}

	\subsubsection{The Complexity of A-HLTL Model Checking.}
We first show the \AHLTL model checking problem for the alternation-free fragment with 
only one trajectory quantifier is \comp{NL-complete}.
For example, verification of information leak in speculative execution in sequential programs renders a formula of the form $\forall^4\Atau$, which belongs to the alternation-free fragment (more details in Section~\ref{sec:eval}).

\begin{theorem}
	\label{thm:altfree-single}
	\MC{$\exists^+\Etau$}{} and \MC{$\forall^+\Atau$}{} are \comp{NL-complete}.
\end{theorem}
	
%
	
	We now switch to formulas with alternating trace quantifiers. The significance of the next theorem 
	is that a single trajectory quantifier does not change the complexity of model checking as 
	compared to the classic \HyperLTL verification~\cite{bf18}.
	It is noteworthy to mention that several important classes of 
	formulas belong to this fragment. For example, according to Theorem~\ref{thm:alternating-single} 
	while model checking {\em observational determinism}~\cite{zm03} ($\forall\forall\Etau$), {\em 
	generalized noninference}~\cite{McLean:1994:GeneralTheory} ($\forall\forall\exists\Etau$), and 
	\emph{non-inference}~\cite{cs10} ($\forall\exists\Etau$) with a single initial input are all 
	\comp{coNP-complete}.

	\begin{theorem}
		\label{thm:alternating-single}
		\MC{$\exists(\exists/\forall)^{+}(\Atau/\Etau)$}{k} is \comp{${\Sigma^p_{k}}$-complete} 
		and \MC{$\forall(\forall/\exists)^{+}(\Etau/\Atau)$}{k} is 
		\comp{${\Pi^p_{k}}$-complete} in the size of the Kripke structure.
	\end{theorem}

We now focus on formulas with multiple trajectory quantifiers.
We first show that alternation-free multiple trajectory quantifiers bumps the class of complexity by 
one step in the polynomial hierarchy.
\begin{theorem}
	\label{thm:alternating-multiple}
	\MC{$\exists(\exists/\forall)^{+}\Etau\Etau^+$}{k} is 
	\comp{${\Sigma^p_{k+1}}$-complete} and  
	\MC{$\forall(\forall/\exists)^{+}\Atau\Atau^+$}{k} is 
	\comp{${\Pi^p_{k+1}}$-complete} in the Kripke structure.
\end{theorem}

\begin{theorem}
		\label{thm:alternating-multiple-alternating}
For $k \geq 1$, \MC{$\exists(\exists/\forall)^{+}\Atau^+\Etau^+$}{k} is 
\comp{${\Sigma^p_{k+1}}$-complete} and \linebreak 
\MC{$\forall(\forall/\exists)^{+}\Etau^+\Atau^+$}{k} is 
\comp{${\Pi^p_{k+1}}$-complete} in the size of the Kripke structure.
\end{theorem}

Finally, Theorems~\ref{thm:alternating-single},~\ref{thm:alternating-multiple}, 
and~\ref{thm:alternating-multiple-alternating} imply that the model 
checking problem for acyclic 
Kripke structures and \AHLTL formulas with an arbitrary number of trace quantifier alternation and 
only one trajectory quantifier is in \comp{PSPACE}. 


	 \section{Case Studies and Evaluation}
\label{sec:eval}

We evaluated our algorithm in Section~\ref{sec:bmc} on cases that require single or nested 
trajectories.
The trajectory encoding presented in Section~\ref{sec:bmc} is implemented on top of the 
open-source bounded model checker \HyperQube~\cite{hsb21},
%
and the QBF solver QuABs~\cite{t19}. 
%
%
All experiments are executed on  a MacBook Pro with 2.2GHz processor and 16GB RAM\footnote{\url{https://github.com/TART-MSU/async_hltl_tacas23}}. 


\begin{wrapfigure}{r}{.255\textwidth}
	\vspace{-1cm}
	\scalebox{.7}{
		\input{figs/progLEAKS2}
	}
	\caption{\small Program with nondeterministic sequence of inputs.}
	\label{fig:concleak_2}
	\vspace{5mm}
\end{wrapfigure}
\subsubsection{Non-interference in Concurrent Programs.}
%
%
%
We first consider the programs presented earlier in Figs.~\ref{fig:prog1} and~\ref{fig:prog2} together 
with \AHLTL formulas $\varphi_\propNI$ and $\varphi_\propNIndet$ from 
Section~\ref{sec:intro}.
%
%
%
We receive UNSAT (for the original formula and not its negation), which indicates that 
violations have been spotted.
Indeed, our implementation successfully finds a counterexample with a specific trajectory that prints 
out \code{`acdb'} when the high-security value \code{h} is equal to zero
%
(entries of $\progacdb$ and $\progacdb_\progndet$ in Table~\ref{tab:exp_results}).
%
%
%
Our other experiment is an extension of the example in~\cite{g07} for multiple 
asynchronous channels (see Fig.~\ref{fig:concleak_2}) and the following formula:
$
\varphi_\propODndet = \forall \pi.\forall \pi'. \Atau \tau.\ \Etau \tau'.\ 
\Always\ (\proplow_{\pi,\tau} \leftrightarrow \proplow_{\pi', \tau}) \rightarrow \Always\ 
(\obs_{\pi,\tau'} 
\leftrightarrow \obs_{\pi', \tau'})
$.
%
%
The results for this case are entries of $\progxy$ and $\progxy_\progndet$ in 
Table~\ref{tab:exp_results}. Details of the counterexample can be found 
in Appendix~\ref{appendix:concurrent}.


%

\vspace{-3mm}
\subsubsection{Speculative Information Flow.}

{\em Speculative execution} is a standard optimization technique that allows branch 
prediction by the processor. 
%
%
%
%
{\em Speculative non-interference} (SNI)~\cite{guarnieri2020spectector} requires that two 
executions with the same {\em policy} $\policy$ (i.e., initial configuration) can be observed 
differently in speculative semantics (e.g., a possible branch), if and only if their 
non-speculative semantics with normal condition checks are also observed 
differently; i.e., the following \AHLTL formula:
\begin{align*}
	\varphi_{\propSNI} & =   \underbrace{\forall \pi_1.\forall 
	\pi_2.}_{\text{speculative}}\  \underbrace{\forall \pi_1'.\forall 
	\pi_2'}_{\text{nonspeculative}}. \ \Atau\tau. 	
	\Big(\Always(\obs_{\pi_1, \tau} \leftrightarrow \obs_{\pi_2, \tau}) \, \wedge    \\ 
	& \hspace{-3mm}(\policy_{\pi_1, \tau} \leftrightarrow \policy_{\pi_2, \tau})  \land  
(\policy_{\pi_1, \tau} \leftrightarrow \policy_{\pi'_1, \tau}) \wedge (\policy_{\pi_2, 
	\tau} \leftrightarrow \policy_{\pi'_2, \tau})\Big) \rightarrow \Always \big(\obs_{\pi_1', \tau} 
	\leftrightarrow 
	\obs_{\pi_2', \tau}\big)
\end{align*}
%
where $\obs$ is the memory footprint, traces $\pi_1$ and $\pi_2$ range 
over the (nonspeculative) \code{C} code and traces $\pi'_1$ and $\pi'_2$ range over the 
corresponding (speculative) assembly code.
We evaluate SNI on the translation from a \code{C} program in Fig.~\ref{fig:se_ccode} 
, where \code{y} is the input policy 
$\policy$ and multiple 
versions of \code{x86} assembly code~\cite{guarnieri2020spectector} 
(details in Appendix~\ref{appendix:speculative}).
%
The results of model checking speculative execution are in Table~\ref{tab:exp_results}
(see entries from $\progse_{V1}$ to $\progse_{V7}$).
Additional versions from $\progse_{V3}$ to  $\progse_{V7}$ are under different 
compilation options. Our method correctly identify all the insecure and secure ones as 
stated in~\cite{guarnieri2020spectector}.

\subsubsection{Compiler Optimization Security.}
Secure compiler optimization~\cite{namjoshi2020witnessing} aims at preserving input-output 
behaviors of a {\em source} program (original implementation) and a {\em target} program (after 
applying optimization), including security policies.
%
We investigate the following optimization strategies: Dead Branch Elimination (DBE), Loop 
Peeling (LP), and Expression Flattening (EF).
To verify a secure optimization, we consider two scenarios: (1) one single I/O event (one trajectory, 
similar to~\cite{bcbfs21}), and (2) a sequences of I/O events (two trajectories): 
\begin{align*}
\varphi_\opt & = \forall \pi.\forall \pi'.  \Etau \tau.\ 
(\propin_{\pi,\tau} \leftrightarrow \propin_{\pi', \tau}) \rightarrow \Always\ (\propout_{\pi,\tau} 
\leftrightarrow \propout_{\pi', \tau})\\
\varphi_\optndet & =  \forall \pi.\forall \pi'. \Atau \tau.\ \Etau \tau'. \Always\ (\propin_{\pi,\tau} 
\leftrightarrow \propin_{\pi', \tau}) \rightarrow \Always\ 
(\propout_{\pi,\tau'} \leftrightarrow \propout_{\pi', \tau'}), 
\end{align*}
where \propin ~ is the set of inputs and \propout ~ is the set of outputs.
%
%
%
Table~\ref{tab:exp_results} (cases $\progdbe$ -- $\progeflp_\progndet$) shows the 
verification results of each optimization strategy and different combination of the strategies 
(details in Appendix~\ref{appendix:compileropt}).

\subsubsection{Cache-Based Timing Attacks.}
Asynchrony also leads to attacks when system executions are confined to a single CPU and its cache~\cite{stefan2013eliminating}.  
A cache-based timing attack happens when an attacker is able to guess the values of high-security variables when cache operations (i.e., evict, fetch) influence the scheduling of different threads.  
Our case study is inspired by the cache-based timing attack example 
in~\cite{stefan2013eliminating} and we use the formula of observational 
determinism $\varphi_\propODndet$ introduced earlier in this section to 
find the potential attacks (see cases of $\progcta$ and $\progcta_\progndet$ in Table~\ref{tab:exp_results})
The details of the case study is discussed in 
Appendix~\ref{appendix:chcheattack}.

\subsection{Analysis of Experimental Results}

Table~\ref{tab:exp_results} presents the diameter of the transition relation, length of trajectories 
$m$, state spaces, and the number of trajectory variables. 
We also present the total solving time of our algorithm as well as the break down: generating models 
(\code{genQBF}), building trajectory encodings (\code{buildTr}), and final QBF solving 
(\code{solveQBF}).
Our two most complex cases are concurrent leak ($\progxy_\progndet$) and loop peeling 
($\proglp_\progndet$).
For concurrent leak, it is because there are three threads with many interleavings (i.e., asynchronous 
composition), takes longer time to build.
For loop peeling, although there is no need to consider interleavings except for the nondeterministic 
inputs; however, the diameters of traces ($D_{\krip_1}$, $D_{\krip_2}$) are longer than 
other cases, which 
makes the length and size of trajectory variables (i.e., $m$ and $|T|$) grow and increases the total 
solving time.
\begin{wraptable}{r}{.52\textwidth}
	\vspace*{-6mm}
	\small
	\renewcommand{\arraystretch}{1.2}
	\centering 
	\scalebox{.7}{
		
		\begin{tabular}{|l || r || c || r |} 
			\cline{2-4}
			\multicolumn{1}{c }{ } & \multicolumn{1}{|c||}{MCHyper~\cite{bcbfs21} } & 
			\multicolumn{2}{c|}{This paper} \\
			\hline
			\thead[c]{\small \bf Case}   
			& {\bf Total[s]}
			&  \code{genQBF}/ \code{buildTr}/ \code{solveQBF}[s]&  {\bf Total}[s]  \\ [0.5ex]
			\hline
			$\progdbe$  &  {\bf 0.8} & 0.9 / 0.07 /  0.01 & {\bf 0.98} \\ 
			\hline
			$\proglp$  & {\bf 365.9} & 1.37  / 1.40 /  1.13 & {\bf 3.90}\\ 
			\hline
			$\progeflp$    & {\bf 1315.2} & 5.11 / 8.12 / 9.35 & {\bf 22.58} \\ 
			\hline
		\end{tabular}

	}
	\caption{Comparison of model checking compiler optimization with~\cite{bcbfs21}. 
	}
	\label{tab:exp_comp_results}	
	\vspace*{-8mm}
\end{wraptable}
Our encoding is able to handle a variety of cases with one or more trajectories, depending on 
whether multiple sources of non-determinism is present.  
To see efficiency, we compare the solving time for cases of compiler optimization with one trajectory 
with the results in~\cite{bcbfs21}. This method reduces \AHLTL model checking to \HyperLTL model 
checking for limited fragments and utilizes the model checker \code{MCHyper}.
On the other hand, in this paper, we directly handle the asynchrony by trajectory encoding 
presented in Section~\ref{sec:bmc}. 
Table~\ref{tab:exp_comp_results} shows our algorithm considerably outperforms the approach 
in~\cite{bcbfs21} in larger cases. 

\begin{table*}[t!]
	\renewcommand{\arraystretch}{1.1}
	\centering 
	\scalebox{.746}{	
		
		\begin{tabular}{|l | l || c c c c c c c || r | r | r || r |} 
			\cline{3-13}
			\multicolumn{2}{c}{ } & \multicolumn{7}{|c||}{(model checking spec and data)} & \multicolumn{4}{c|}{(time took for solving)} \\
			\hline
			\thead[c]{\small \bf Models}  & \thead[c]{\small $\varphi$} & $D_{\krip_1}$ & $D_{\krip_2}$ & $m$  & $ 
			|S_{\krip_1}| $ & $ |S_{\krip_2}| $ & $|T|$ & ~QBF~ &  \code{genQBF}[s] &  \code{buildTr}[s] & 
			\code{solveQBF}[s]&  {\bf Total}[s]  \\ [0.5ex]
			\hline 
			\hline
			$\progacdb$ & $\varphi_\propNI$& 6 & 6  & 12 & 109 & 109 & 1378 & \UNSAT & 2.80 &  0.32 & 0.23 & {\bf 3.35}    \\
			\hline
			$\progacdb_\progndet$ & $\varphi_\propNIndet$& 8 & 8 & 16 & 696  & 696  & 2754  & \UNSAT   & 7.74 &   2.54& 3.73& {\bf 14.01}  \\
			\hline\hline
			$\progxy$ & $\varphi_\propOD$ & 11 & 11 & 22 & 597 & 597 &  6118& \UNSAT & 14.85& 7.10& 8.29 & {\bf 30.24 }  \\
			\hline
			$\progxy_\progndet$ & $\varphi_\propODndet$ & 18 & 18 & 36 & 2988 & 2988 &  22274 & \UNSAT & 127.09 & 53.14 & 731.48 &{\bf  911.72 }    \\
			\hline\hline
			$\progse_{V1}$ & $\varphi_{\propSNI}$ & 3 & 6 & 9 & 132 & 340 & 1112 & \UNSAT & 7.45&  1.72&  3.07 & {\bf 12.24}\\  
			\hline
			$\progse_{V2}$ & $\varphi_{\propSNI}$  & 3 & 6 & 9 &  144 & 168&  1112 & \SAT & 5.61&  1.28 &   2.44  & {\bf  9.33} \\    
			\hline
			$\progse_{V3}$ & $\varphi_{\propSNI}$ & 3 & 6 & 9 & 87 & 340 &636 & \UNSAT &  7.30&  1.68&  2.97& {\bf 11.95}\\
			\hline
			$\progse_{V4}$ & $\varphi_{\propSNI}$ & 3 & 6  & 9 & 93& 340 & 636 & \UNSAT &  7.37 &  1.71 &  4.50  & {\bf 13.58 }\\
			\hline
			$\progse_{V5}$ & $\varphi_{\propSNI}$ & 3 & 6 & 9 & 132 & 168  &636 & \SAT & 6.23 &  1.23 &   3.48 & {\bf 10.94} \\ 
			\hline
			$\progse_{V6}$ & $\varphi_{\propSNI}$ &  3 & 7 & 10  &   132 & 340 & 766 & \UNSAT & 7.47&  1.82&  3.26   & {\bf 12.55}\\ 
			\hline
			$\progse_{V7}$ & $\varphi_{\propSNI}$ &  2& 5 &  7  & 144 & 168 & 352 & \SAT & 5.83 &  1.28 & 2.58 & {\bf 9.69}\\ 
			\hline 
			\hline
			$\progdbe$  & $\varphi_\opt$  &  4 & 4  & 8 & 8 & 6 & 546 & \SAT & 0.9 & 0.07 &  0.01 & {\bf 0.98} \\ 
			\hline
			$\progdbe_\progndet$  & $\varphi_\optndet$  &  13 & 13  & 26 & 82& 72& 9414& \SAT & 1.60& 0.56&  9.61 & {\bf 11.77}\\ 
			\hline
			$\progdbe_\progndet$ {\scriptsize w/ bugs}  & $\varphi_\optndet$ &  13 & 13  & 26 & 82& 72& 9414& \UNSAT & 1.36 &0.49&  2.05 & {\bf 3.90}\\
			\hline
			$\proglp$  & $\varphi_\opt$  &  22 & 22  & 44 & 80 & 76 & 3870 & \SAT & 1.37  & 1.40 &  1.13 & {\bf 3.90}\\ 
			\hline
			$\proglp_\progndet$    & $\varphi_\optndet$  &  17 & 17  & 34 & 558& 811& 19110& \SAT & 7.37&3.86&  48.15&  {\bf 59.38} \\ 
			\hline
			$\proglp_\progndet$ {\scriptsize w/ loops}  & $\varphi_\optndet$  & 33 & 35 & 68  & 757  & 1591  & 128114 & \SAT  & 30.52  & 34.99 & 4165.54 & {\bf 4231.05} \\
			\hline
			$\proglp_\progndet$ {\scriptsize w/ bugs}  & $\varphi_\optndet$ &  17 & 17  & 34 & 558 & 661& 19110& \UNSAT & 6.51&3.60&  20.75 & {\bf 30.86}\\ 
			\hline
			$\progeflp$  & $\varphi_\opt$  &  32 & 32  & 64 & 80 & 248 & 108290 & \SAT & 5.11 & 8.12 & 9.35 & {\bf 22.58} \\ 
			\hline
			$\progeflp_\progndet$  & $\varphi_\optndet$  &  18 & 22  & 40 & 582 & 1729 & 28986 & \SAT & 15.92 & 8.90 &  135.48 & {\bf 160.30} \\ 
			\hline
			$\progeflp_\progndet$ {\scriptsize w/ loops}  & $\varphi_\optndet$  &  33 & 45  & 78 & 295 & 1996 & 178894 & \SAT & 36.98 & 62.89 &  121.60 & {\bf 221.47} \\ 
			\hline
			\hline
			$\progcta$  & $\varphi_\propOD$  & 13&  13&  26 & 48  & 48  & 9414  & \UNSAT  &  1.49 &  0.53 & 0.38 & {\bf 2.40 } \\ 
			\hline
			$\progcta_\progndet$  & $\varphi_\propODndet$  & 58  & 58  & 16 & 16 & 32 & 16258  &  \UNSAT  &  1.95 & 1.33 &  1.02 & {\bf 4.30  } \\ 
			\hline
			$\progcta_\progndet$ {\scriptsize w/ loops} & $\varphi_\propODndet$  &  35 &  35 &  70 & 88 & 88  & 139302 & \UNSAT & 5.50  & 27.65& 125.92 & {\bf 159.07 } \\ 
			\hline
		\end{tabular}

	}
	\vspace{2mm}
	\caption{Case studies break down for Kripke structures: $\krip_1, \krip_2$ (all case studies have 
		two, e.g.,one for high-level and one for assembly code), formula: 
		$\varphi$, diameter: $D$, state space: $|S|$, trajectory depth: $m$, and size of trajectory 
		variables: $|T|$.}
	\label{tab:exp_results}	
	\vspace{-9mm}
\end{table*}

\section{Conclusion and Future Work}
\label{sec:concl}
In this paper, we focused on the problem of \AHLTL model checking for {\em terminating} programs.
 %
%
We generalized \AHLTL to allow nested {\em trajectory} quantification, where a trajectory determines how different traces may advance and stutter.
%
%
We rigorously analyzed the complexity of  \AHLTL model checking for acyclic Kripke structures.
The complexity grows in the polynomial hierarchy with the number of 
quantifier alternations, and, it is either 
aligned with that of \HyperLTL or is one step higher in the polynomial hierarchy.
We also proposed a BMC algorithm for \AHLTL based on QBF-solving and 
reported successful experimental results on verification of information flow security in concurrent 
programs, speculative execution, compiler optimization, and cache-based timing attacks.

Asynchronous hyperproperties enable logic-based 
verification for software programs. 
%
Thus, future work includes developing different abstraction techniques 
such as predicate abstraction, abstraction-refinement, etc, to develop 
software model checking techniques.
We also believe developing synthesis techniques for \AHLTL creates opportunities to automatically 
generate secure programs and assist in areas such as secure compilation.

\section*{Acknowledgment}

\Thanks

	\bibliographystyle{abbrv}
	\bibliography{bibliography}
\newpage
	\appendix
	\section{Detailed Proofs}
\label{sec:proofs}

\subsection*{Proof of Theorem~\ref{thm:qbf-mc-unroll-new}}

From our construction in Section~\ref{sec:bmc}, the following lemma follows.

\begin{lemma}
	\label{thm:qbf-mc}
	Let $\varphi$ be an \AHLTL formula with at most one trajectory
	quantifier alternation and $k$ and $m$ be unrolling bounds. Then,
	\begin{compactenum}
		\item If $\QBF{\krip,\varphi}_{k,m}^\HPES$ is satisfiable, then
		$\krip\models\varphi$.
		\item If $\QBF{\krip,\varphi}_{k,m}^\HOPT$ is unsatisfiable, then
		$\krip\not\models\varphi$.
	\end{compactenum}
\end{lemma}

Let $K$ be the maximum length of any path in any Kripke structure.
It is easy to see that in all cases, after at most
$K*|\Paths|*|\Trajs|$ steps, all paths have halted according to all
trajectories because at every step there is always some trajectory
moving some non-halted path.
Since the halting optimistic and the pessimistic semantics only differ
when the paths do not halt after the unrolling limit consider, the
following result holds.

\begin{lemma}
	\label{thm:qbf-mc-unroll}
	Let $\varphi$ be an \AHLTL formula with at most one trajectory
	quantifier alternation let $K$ be the maximum depth of a Kripke
	structure and let $M=K*|\Paths(\varphi)|*|\Trajs(\varphi)|$.
	Then,
	\(
	\QBF{\krip,\varphi}_{K,M}^\HPES=\QBF{\krip,\varphi}_{K,M}^\HOPT.
	\)
\end{lemma}


Finally, Lemmas~\ref{thm:qbf-mc} and~\ref{thm:qbf-mc-unroll} imply
that after the unrolling bound $k*|\Paths(\varphi)|*|\Trajs(\varphi)|$
both the halting optimistic and pessimistic give the correct answer to
the model checking problem.

\subsection*{Proof of Theorem~\ref{thm:altfree-single}}		For the upper bound, we consider 
the 
case that the \AHLTL formula is 
existential, i.e., it is of the form:
$$
\exists \pi_1 \ldots \exists \pi_k . \Etau \tau. \, \varphi,
$$
where $\varphi$ does not contain any trace quantifiers.
For the case that the formula is universal, i.e., it is of the form:
$$
\forall \pi_1 \ldots \forall \pi_k . \Atau \tau. \, \varphi,
$$
we check the formula $\exists \pi_1 \ldots \exists \pi_k . \Etau \tau. \, \neg \varphi$
and report the complemented result.

The algorithm for the upper bound works as follows. Since the Kripke structure is acyclic, the length of the traces is bounded by the number of states of the Kripke structure.
We can, therefore, nondeterministically guess the witness to traces $\pi_1 \cdots \pi_k$ and trajectory $\tau$ that satisfy the inner \LTL formula $\varphi$ using a counter per trace (with a logarithmic number of bits in the number of states of $\krip$) and $k$ bits for the trajectory. 
Observe that $\tau$ merely prescribes how the traces advance. That means one can obtain traces $\sigma_1 \cdots \sigma_k$ from the witnesses to $\pi_1 \cdots \pi_k$ that advance synchronously (i.e., all traces advance in a lockstep manner), where the length of traces is dictated by the guessed witness to $\tau$.
Since verifying the correctness of $\varphi$ on these traces (that form a tree-shaped graph)  can be done in logarithmic time~\cite{bf18}, the upper bound remains in \comp{NL}.
We emphasize that the number of counters and extra $k$ bits are in the size of the formula which is assumed to be a constant, as our complexity analysis is in the size of the input Kripke structure.

The lower bound follows from the \comp{NL-hardness} of standard \HyperLTL model checking for acyclic graphs~\cite{bf18}.\qed

\subsection*{Proof of Theorem~\ref{thm:alternating-single}}

We show membership in \comp{${\Sigma^p_{k}}$} and 
\comp{${\Pi^p_{k}}$}, respectively, by induction over $k$.
We begin with the base case ($k=1$), that is, the fragment $\exists^+\Atau$. By 
nondeterministically guessing the 
witnesses to the existential trace quantifiers, according to Theorem~\ref{thm:altfree-single}, the 
model checking problem for only a universal trajectory quantifier is solvable in polynomial time.
That means for $k=1$, \MC{$\exists^+\Atau$}{} is in \comp{NP = $\Sigma^p_1$}.
Dually, for $k=1$, \MC{$\forall^+\Etau$}{} is in \comp{coNP = $\Pi^p_1$}.

For the inductive step, let us first focus on decision problem 
\MC{$\exists(\exists/\forall)^+\Atau$}{k}.
Since the Kripke structure is acyclic, the length of the traces is bounded by the number of 
states.
We can, thus, nondeterministically guess the existentially quantified traces in polynomial time and 
then verify the correctness of the guess, by the induction hypothesis, in \comp{${\Pi^p_{k-1}}$}. 
Hence, the model checking problem for $k$ alternations is in \comp{${\Sigma^p_{k}}$}. 
Likewise, for the decision problem  \MC{$\forall(\forall/\exists)^+\Etau$}{k}, we universally guess the 
universal quantified traces in polynomial time and verify the 
correctness of the guess, by the induction hypothesis, in \comp{${\Sigma^p_{k}}$}.
Hence, the problem of determining $\krip \models \varphi$ for $k$ alternations in $\varphi$ is in 
\comp{${\Pi^p_{k}}$}. 

For the lower bound, we show that \MC{$\exists(\exists/\forall)^{+}\Atau$}{k} and 
\linebreak \MC{$\forall(\forall/\exists)^{+}\Etau$}{k} are \comp{${\Sigma^p_{k}}$-hard} and 
\comp{${\Pi^p_{k}}$-hard}, respectively, via a reduction from the {\em quantified Boolean 
	formula} (QBF) satisfiability problem~\cite{gj79}:

\begin{quote}
	
	{\em Given is a set of Boolean variables, $\{x_1, x_2, \dots, x_n\}$, and a 
		quantified Boolean formula
		$$
		y=\quant_1 x_1.\quant_2 x_2\dots\quant_{n-1} x_{n-1}.\quant_n x_n.(y_1 \, 
		\wedge \, y_2 \, \wedge \dots \wedge \, y_m)
		$$
		where each $\quant_i \in \{\forall, \exists\}$ ($i \in [1, n]$) and each clause 
		$y_j$ ($j \in [1, m]$) is a disjunction of three literals (3CNF). Is $y$ 
		true?}
	
\end{quote}
If $y$ is restricted to at most $k$ alternations of quantifiers, then QBF 
satisfiability is complete for \comp{${\Sigma^p_{k+1}}$} if $\quant_1 = 
\exists$, and for \comp{${\Pi^p_{k}}$} if $\quant_1 = \forall$. We note 
that in the given instance of the QBF problem:

\begin{itemize}
	
	\item The clauses may have more than three literals, but three is sufficient of 
	our purpose;
	
	\item The inner Boolean formula has to be in conjunctive normal form in order 
	for our reduction to work;
	
	\item Without loss of generality, the variables in the literals of the same 
	clause are different (this can be achieved by a simple pre-processing of the 
	formula), and
	
	\item If the formula has $k$ alternations, then it has $k+1$ alternation {\em 
		depths}. For example, formula
	$$\forall x_1.\exists x_2. (x_1 \vee \neg x_2)$$ 
	has one alternation, but two alternation depths: one for $\forall x_1$ and the 
	second for $\exists x_2$. By $d(x_i)$, we mean the alternation depth of Boolean 
	variable $x_i$.
	
\end{itemize}

\begin{figure}[t]
	\centering
	\includegraphics[scale=1]{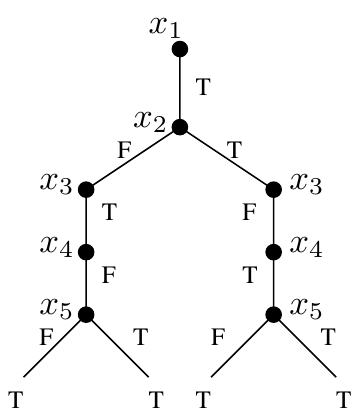}
	\caption{Model for the QBF $y = \exists x_1.\forall 
		x_2.\exists x_3.\exists x_4.\forall x_5.(x_1 \vee \neg x_2 \vee x_3) \wedge 
		(\neg x_1 \vee x_2 \vee \neg x_4) \wedge (\neg x_3 \vee x_4 \vee \neg x_5) 
		\wedge (x_1 \vee x_4 \vee x_5)$.}
	\label{fig:qbf}
\end{figure}

\begin{figure}[t]
	\centering
	\includegraphics[scale=.8]{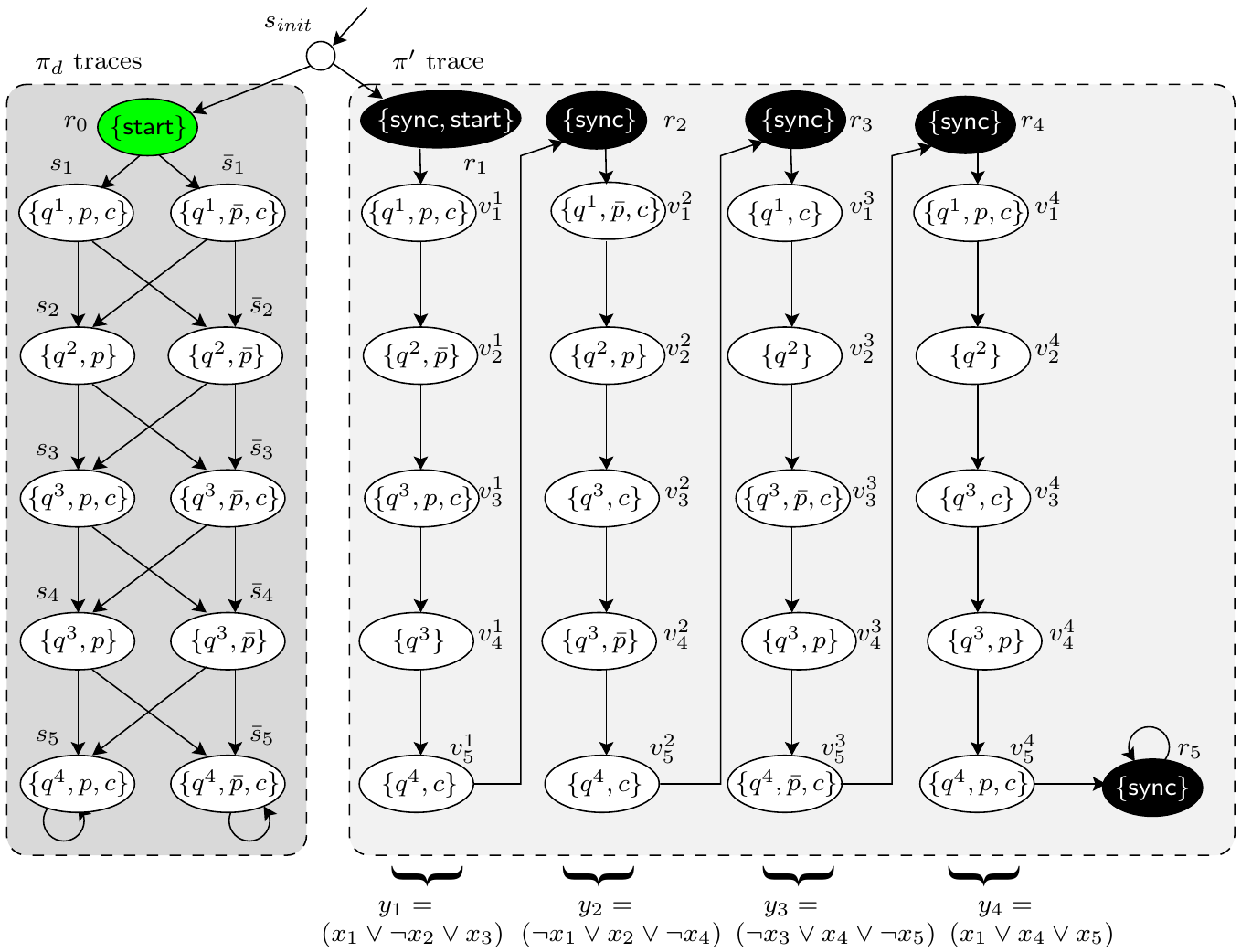}
	\caption{Mapping quantified Boolean formula $y = \exists x_1.\forall 
		x_2.\exists x_3.\exists x_4.\forall x_5.(x_1 \vee \neg x_2 \vee x_3) \wedge 
		(\neg x_1 \vee x_2 \vee \neg x_4) \wedge (\neg x_3 \vee x_4 \vee \neg x_5) 
		\wedge (x_1 \vee x_4 \vee x_5)$ to an instance of 
		\MC{$\exists(\exists/\forall)^k\Atau$}.}
	\label{fig:singletraj}
\end{figure}

We now present a mapping from an arbitrary instance of QBF with $k$ 
alternations and where $\quant_1 = \exists$ to the model checking problem of an acyclic 
Kripke structure and a \AHLTL formula with $k$ trace quantifier alternations and one innermost universal trajectory quantifier.
Then, we show that the Kripke structure satisfies the \AHLTL formula if and only if 
the answer to the QBF problem is affirmative.
Figures~\ref{fig:qbf} and~\ref{fig:singletraj} show an example.

\noindent \textbf{Kripke structure $\krip = \ktuple$: } 

\begin{itemize}
	
	\item {\em (Atomic propositions $\AP$)} For each alternation depth $d \in [1, 
	k+1]$, we include an atomic proposition $q^d$. We furthermore include five 
	atomic propositions: $p$ is used to force 
	clauses to become true if a positive literal appears in a clause; proposition $\bar{p}$ is used to 
	force clauses to become true if a negative literal appears in a clause in our 
	reduction; $\start$ marks the beginning of the gadget of states that represent the Boolean 
	variables in the QBF instance; $\sync$ is used to mark the beginning a chain of states that 
	represent a clause, and $c$ is used to enforce lock-step synchronization of Boolean variables in 
	their respective clauses. Thus,
	$$
	\AP = \big\{c, p, \bar{p}, \start, \sync\big\} \; \cup \; \big\{ q^d \mid d \in [1, 
	k+1]\big\}.
	$$
	
	\item {\em (Set of states $\States$)}  We now identify the members of $\States$:
	
	\begin{itemize}
		
		\item First, we include an initial state $\state_\init$  labeled by $\start$. 
		
		\item For each Boolean variable $x_i$, where $i \in [1, n]$, we include three 
		states $s_i$ and $\bar{s}_i$. Each state $s_i$ (respectively, 
		$\bar{s}_i$) is labeled by $p$ and $q^{d(x_i)}$ (respectively, $\bar{p}$ and
		$q^{d(x_i)}$).
		
		\item For each clause $y_j$, where $j \in [1, m]$, we include a state 
		$r_j$, labeled by proposition $\sync$. We also include state $r_{m+1}$ labeled by $\sync$ that 
		marks the end of chain of states that represent the literals in the clauses.
		
		\item For each clause $y_j$, where $j \in [1, m]$, we introduce the 
		following $n$ states: 
		$$
		\Big\{v^j_i \mid i \in [1, n]\Big\}.
		$$
		Each state $v^j_i$ is labeled with propositions $q^{d(x_i)}$, and with $p$ if 
		$x_i$ is a literal in $y_j$, or with $\bar{p}$ if $\neg x_i$ is a literal in $y_j$.
		
		\item Finally, we label states $s_i$, $\bar{s}_i$, and $v_i^j$ by proposition $c$, if $i$ is odd, for 
		all $i \in [1, n]$.
		
	\end{itemize}
	Thus,
	\begin{align*}
		S = & \big\{s_\init \big\} \, \cup \, \big\{r_j \mid j \in [0, m+1]\big\}  \; 
		\cup \; \big\{v^j_i, s_i, \bar{s_i}, \mid i \in [1, n] \wedge 
		j \in [1,m]\big\}.
	\end{align*}
	
	\item {\em (Transition relation $\trans$)} We now identify the members of 
	$\trans$:
	\begin{align*}
		\trans = & \big\{(\state_\init, r_0), (\state_\init, r_1)\} \; \cup \; \big\{(r_0, s_1), (r_0, \bar{s}_1), 
		(v_n^m, r_{m+1})\big\} \; \cup\\
		& \big\{(v^j_i, v^j_{i+1}) \mid i \in [1, n) \, \wedge \, j \in [1, m] \big\} \; 
		\cup\\
		& \big\{(s_i, {s}_{i+1}), (\bar{s}_i, {s}_{i+1}), ({s}_i, \bar{s}_{i+1}), (\bar{s}_i, \bar{s}_{i+1}) \mid i 
		\in [1, n) \big\} \; \cup \\
		& \big\{(s_n, s_n), (\bar{s}_n, \bar{s}_n), (r_{m+1}, r_{m+1})\big\}.
	\end{align*}
	
\end{itemize}


\noindent \textbf{A-HLTL formula: } The \AHLTL formula in our mapping is 
the following:

{
\begin{tikzpicture}
	\node (a1) at (-.5, 1) {$\mathcal{Q}\pi'$};
	\draw[-] (-2.1,.75) -- (.5,.75);
	\draw[->] (-2.1,.75) -- (-2.1, .2);
	\draw[->] (.5,.75) -- (.5,.2);
\node (a) at (-1, 0) {$\varphi_{\map} = \quant_{1} \pi_{1}.\quant_2 \pi_2\cdots \quant_n \pi_n 
.\Atau\tau. $};
\end{tikzpicture}

$$\Bigg( \bigwedge_{d \in \{ i \mid \quant_i = \forall\}} \underbrace{\F(
\start_{\pi_d,\tau} \wedge \neg \sync_{\pi_d,\tau})}_{\psi_1} \, \wedge \, \underbrace{\F 
(\start_{\pi',\tau} \wedge \sync_{\pi',\tau})}_{\psi_2}\Bigg)
$$

$$
\longrightarrow 
$$

\vspace{-3mm}
\[
\begin{pmatrix}
	\Bigg[  
	\bigwedge_{d \in \{i \mid \quant_i = \exists\}} \underbrace{\F (\start_{\pi_d,\tau} \wedge \neg 
	\sync_{\pi_d,\tau})}_{\psi_3}\Bigg] \; \wedge\\
	\Always\Bigg[ \underbrace{\bigwedge_{d \in [1, n]}\Big(\sync_{\pi',\tau}  \wedge 
	\big(\sync_{\pi', \tau} \Until (\neg 
	\sync_{\pi', \tau} 	\wedge 	(c_{\pi_d,\tau} \Iff c_{\pi',\tau})) \Until 
	\sync_{\pi',\tau}\big)\Big)}_{\psi_4}\\
\longrightarrow \\
	\bigvee_{d \in [1,n]} \underbrace{\F \Big(\big(q^{d}_{\pi_d,\tau} 
	\leftrightarrow q_{\pi',\tau}^d\big) \wedge \big((p_{\pi',\tau} \wedge p_{\pi_d,\tau}) \vee 
	(\bar{p}_{\pi',\tau} \wedge \bar{p}_{\pi_d,\tau})\big)\Big)}_{\psi_5} \Bigg] 
\end{pmatrix}
\]
}
where for each $i \in [1, n]$, $\quant_i$ is the same type of quantifier as in the input QBF 
instance.
Observe that there is only one way to instantiate trace $\pi'$, namely the path that starts $\state_{\init}$ 
and ends in $r_5$.
Thus, we may insert quantifier $\mathcal{Q}\pi'$ in any place such that it does not change the 
number of alternations of  $\quant_{1} \pi_{1}.\quant_2 \pi_2\cdots \quant_n \pi_n 
.\Atau\tau$.
Quantifier $\mathcal{Q}$ can be either $\forall$ or $\exists$. If $\mathcal{Q} =\forall$, then \AHLTL 
formula is as $\varphi_\map$ above.
If $\mathcal{Q} =\exists$, then in $\varphi_\map$ the sub-formula  $\psi_2$ would have to appear as 
a conjunct on the right side of the first implication.

%
Since the input QBF has $k$ alternations and the resulting \AHLTL formulas has $k+1$ alternations. 
Intuitively, this formula expresses the following. First, we limit the universal trajectory to ones 
that only align the states of the gadgets for Boolean variables $x_1 \cdots x_n$ and the states of the 
gadgets for  clauses $y_1 \cdots y_m$  as aligned and they advance in a lock-step manner.
Let us explain the purpose of each sub-formula in $\varphi_{\map}$:

\begin{itemize}
	\item Sub-formulas $\psi_1$ (for universal) and $\psi_3$ (for existential traces) ensure that 
	$\pi_d$ traces that range over the left substructure start from state $r_0$.
	
	\item Sub-formula $\psi_2$ moves the unique trace $\pi'$ to state $r_j$, where $j \in [1,m]$. As 
	mentioned earlier if 
	$\mathcal{Q} = \exists$, then $\psi_2$ will appear as a conjunction with $\psi_3$.
	
	\item Sub-formula $\psi_4$ filters the universally quantified trajectory $\tau$ by allowing only 
	those 
	that (1) when reaching an $r_j$ state, where $\sync$ holds; (2) move traces $\pi_d$ and $\pi'$ in 
	lock-step (i.e., $(c_{\pi_d,\tau} \Iff c_{\pi',\tau})$), where $\pi'$ is not in a $\sync$ state, and (3) 
	until it reaches the last $\sync$ state.
	
	\item Sub-formula $\psi_5$ ensures that each instance of $\tau$ aligns with at least one of the 
	clauses in the input QBF formula and evaluates that clause to true (either $p$ or $\bar{p}$ agree 
	with each other in $\pi_d$ and $\pi'$).  
\end{itemize}

Also, formula $\varphi_\mathsf{map}$ forces $\tau$ to synchronize $\pi_d$ at 
state $r_0$ with $\pi'$ at one of the states 
labeled by $\sync$ (i.e., the beginning of a clause). Then, sub-formula 
$c_{\pi_d,\tau} \Iff c_{\pi',\tau}$ 
ensures that the trajectory advances $\pi$ and 
$\pi'$ in lock-step to determine whether the clause evaluates to true. Now, for all the trajectories 
that meet these conditions, if there exists a state where either $p$ or $\bar{p}$ in $\pi_d$ eventually 
matches its counterpart position in $\pi'$, then clause is satisfied. The matching positions identify 
the assignments of Boolean variables in the corresponding clauses that make the QBF instance 
true. 


We now show that the given quantified Boolean formula is $\mathit{true}$ if and 
only if the Kripke structure obtained by our mapping satisfies the \AHLTL
formula $\varphi_\map$. 

\begin{description}
	
	\item[($\Rightarrow$)] Suppose that $y$ is true. Then, there is an instantiation 
	of existentially quantified trace variables for each value of universally quantified 
	variables and the trajectory quantifier, such that each clause $y_j$, where $j \in [1, m]$ becomes true (see 
	Figs.~\ref{fig:qbf} and~\ref{fig:singletraj} for an example). We now use these instantiations to 
	instantiate each $\exists \pi_{x_d}$ in \AHLTL formula $\varphi_\map$, where 
	$d \in \{i \mid \quant_i = \exists\}$ as follows. First, notice that $\pi'$ can only be instantiated by 
	the trace that reaches state $r_1$. Now, for each existentially quantified 
	variable $x_i$, where $i \in [1, n]$, in depth $d \in [1,k+1]$, if $x_i = 
	\tru$, we instantiate $\pi_d$ with a trace that includes state $s_i$. 
	Otherwise, the trace will include state $\bar{s}_i$. We now show that this 
	trace instantiation evaluates formula $\varphi_\map$ to true. 
	Observe that 
	the left side of the implication in the formula is basically filtering the non-legitimate trajectories and allows only those  where traces $\pi_d$ can synchronize with trace $\pi'$.
	Since each 
	$y_j$ is true, for any instantiation of universal quantifiers, there is at 
	least one literal in $y_j$ that is true. If this literal is of the form $x_i$, 
	then we have $x_i = \tru$ and trace $\pi_d$ will include $s_i$, which is 
	labeled by $p$ and $q^d$. Hence, the values of $p$ (respectively, $q^d$), in 
	both $\pi_d$ and $\pi'$ instantiated by trace
	$$
	\state_{\init}r_1 \cdots r_jv^j_1\cdots v^j_n\cdots r_{m+1}^\omega
	$$
	are eventually equal. If the literal in $y_j$ is of the form $\neg x_i$, 
	then $x_i = \fals$ and, hence, some trace $\pi_d$ will include $\bar{s}_i$. 
	Again, the values of $\bar{p}$ (respectively, $q^d$), in both $\pi_d$ and 
	$\pi'$ are eventually equal. Finally, since all clauses are true, all traces 
	$\pi'$ reach a state where the right side of the implication becomes true.
	
	\item[($\Leftarrow$)] Suppose our mapped Kripke structure satisfies the
	\AHLTL formula $\varphi_\map$. This means that for each instantiation of  the trajectory $\tau$, 
	since trace $\pi'$ is of the form \linebreak $\state_{\init}r_1 \cdots r_jv^j_1\cdots v^j_n\cdots 
	r_{m+1}^\omega$, then there exists a state $v_i^j$, where the values of $q^d$ and either $p$ or 
	$\bar{p}$ are eventually equal to their counterparts in some trace $\pi_d$. If 
	this trace is existentially quantified and includes $s_i$, then we assign $x_i 
	= \tru$ for the preceding quantifications. If the trace includes $\bar{s}_i$, 
	then $x_i = \fals$. Observe that since in no state $p$ and $\bar{p}$ are 
	simultaneously true and no trace includes both $s_i$ and $\bar{s}_i$, variable 
	$x_i$ will have only one truth value. This way, a model similar to 
	Fig.~\ref{fig:qbf} can be constructed. Similar to the forward 
	direction, it is straightforward to see that this valuation makes every clause 
	$y_j$ of the QBF instance true.
	
\end{description}

\begin{wrapfigure}{r}{.4\textwidth}
	\centering
	\vspace{-25pt}
	\includegraphics[scale=.6]{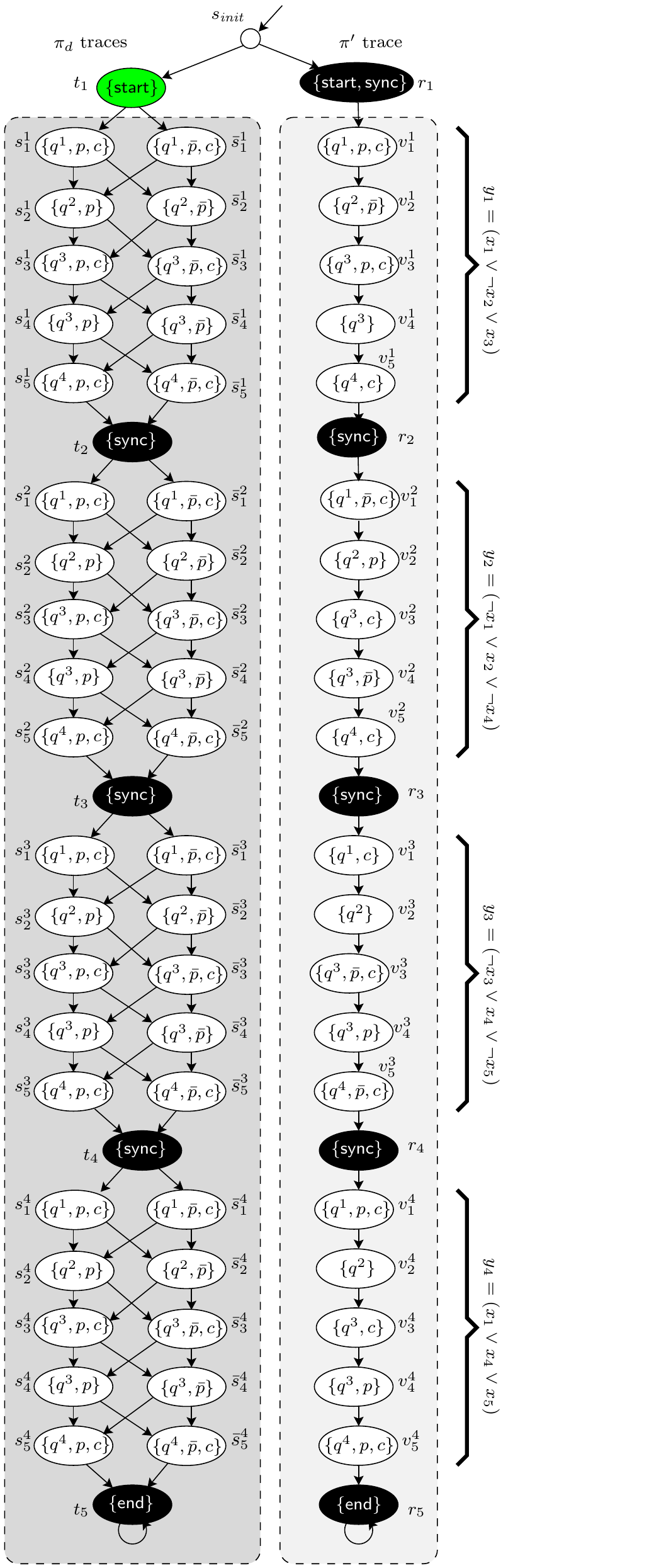}
	\caption{
		Mapping the QBF formula as in Fig.~\ref{fig:singletraj} to an instance of 
		\MC{$\exists(\exists/\forall)^k\Etau\Etau$}.}
	\vspace{-18mm}
	\label{fig:multitraj}
\end{wrapfigure}
In our mapping if $\quant_1 = \exists$, the hardness of model checking for \AHLTL formulas is 
$\Sigma_k^p$. If $\quant_1 = \forall$, then analogously the problem becomes 
\comp{${\Pi^p_{k}}$-hard}.
\qed

%
%

\subsection*{Proof of Theorem~\ref{thm:alternating-multiple}}

We show membership to 
\comp{${\Sigma^p_{k+1}}$} and \comp{${\Pi^p_{k+1}}$}, respectively, by induction over $k$.
We begin with the base case ($k=0$), that is, the fragment $\exists^+\Etau\Etau$. By guessing the 
witnesses to the existential and trajectory quantifiers, we can verify the correctness of the inner \LTL 
formula in polynomial time.
Thus, \MC{$\exists^+\Etau\Etau$}{} is in \comp{NP} and dually  \MC{$\forall^+\Atau\Atau$}{} is in 
\comp{coNP}.
For $k$ quantifier alternations, let us first focus on the decision problem 
\MC{$\exists(\exists/\forall)^{+}\Etau\Etau$}{k}.
Since the Kripke structure is acyclic, the length of the traces is bounded by the number of 
states.
We can, thus, nondeterministically guess the existentially quantified traces in polynomial time 
and 
then verify the correctness of the guess, by the induction hypothesis, in 
\comp{${\Pi^p_{k}}$}. 
Hence, the model checking problem for $k$ is in \comp{${\Sigma^p_{k+1}}$}. 
Likewise, for the decision problem  \MC{$\forall(\forall/\exists)^{+}\Atau\Atau$}{k}, we universally 
guess the universally quantified traces in polynomial time and verify the 
correctness of the guess, by the induction hypothesis, in \comp{${\Sigma^p_{k+1}}$}.
Hence, the problem of determining $\krip \models \varphi$ for $k$ alternations in $\varphi$ is in 
\comp{${\Pi^p_{k+1}}$}.

For the lower bound, similar to the proof of Theorem~\ref{thm:alternating-single}, we show that 
\MC{$\exists(\exists/\forall)^{+}\Etau\Etau$}{k} and 
\MC{$\forall(\forall\exists)^{+}\Atau\Atau$}{k} are \comp{${\Sigma^p_{k+1}}$-hard} and 
\comp{${\Pi^p_{k+1}}$-hard}, respectively, via a reduction from the QBF satisfiability problem (see 
the description of the problem in the proof of Theorem~\ref{thm:alternating-single}).
We now present a mapping from an arbitrary instance of QBF with $k$ alternations to the model 
checking problem of an acyclic Kripke structure and a \AHLTL 
formula with $k$ quantifier alternations and two innermost existential trajectory quantifiers.
Then, we show that the Kripke structure satisfies the \AHLTL formula if and only if 
the answer to the QBF problem is affirmative.
Figure~\ref{fig:multitraj} shows an example of our mapping.\\

\noindent \textbf{Kripke structure $\krip = \ktuple$: } 

\begin{itemize}
	
	\item {\em (Atomic propositions $\AP$)} For each alternation depth $d \in [1, 
	k+1]$, we include an atomic proposition $q^d$. We furthermore include five 
	atomic propositions: $p$ is used to force 
	clauses to become true if a Boolean variable appears in a clause; proposition $\bar{p}$ is 
	used to 
	force clauses to become true if the negation of a Boolean variable appears in a clause in our 
	reduction; $\start$ marks the beginning of the gadget of states that represent the Boolean 
	variables in the QBF instance; $\sync$ is used to mark the beginning a chain of states that 
	represent a clause; $\myend$ marks the terminal states, and $c$ is used to enforce 
	lock-step synchronization of Boolean variables in their respective clauses. Thus,
	$$
	\AP = \big\{c, p, \bar{p}, \start, \sync, \myend \big\} \; \cup \; \big\{ q^d \mid d \in [1, 
	k+1]\big\}.
	$$
	
	\item {\em (Set of states $\States$)}  We now identify the members of $\States$:
	
	\begin{itemize}
		
		\item First, we include an initial state $\state_\init$. 
		
		\item For each Boolean variable $x_i$, where $i \in [1, n]$ and clause $y_j$, where $j \in [1, 
		m]$, we include two states $s_i^j$ and $\bar{s}_i^j$. Each state $s_i^j$ (respectively, 
			$\bar{s}_i^j$) is labeled by $p$ and $q^{d(x_i)}$ (respectively, $\bar{p}$ and
			$q^{d(x_i)}$), and

		\item For each clause $y_j$, where $j \in [1, m]$: 
		
		\begin{itemize}
			\item We include states $r_j$ and $t_j$, labeled by proposition $\sync$. We also include 
			states $r_{m+1}$ and $t_{m+1}$ labeled by $\myend$ that marks the end 
			of chain of states that represent the literals in the clauses.

			\item We introduce the following $n$ states: 
		$$
		\Big\{v^j_i \mid i \in [1, n]\Big\}.
		$$
		Each state $v^j_i$ is labeled with propositions $q^{d(x_i)}$, and with $p$ if 
		$x_i$ is a literal in $y_j$, or with $\bar{p}$ if $\neg x_i$ is a literal in $y_j$.
		
	\end{itemize}
		\item Finally, we label states $s_i^j$, $\bar{s}_i^j$, and $v_i^j$ by proposition $c$, if $i$ is 
		odd, for all $i \in [1, n]$ and $j \in [1, m]$.
		
	\end{itemize}
	Thus,
	\begin{align*}
		S = & \big\{s_\init \big\} \, \cup \, \big\{r_j, t_j \mid j \in [0, m+1]\big\}  \; 
		\cup \; \big\{v^j_i, s_i^j, \bar{s}_i^j \mid i \in [1, n] \wedge j \in [1,m]\big\}.
	\end{align*}
	
	\item {\em (Transition relation $\trans$)} We now identify the members of 
	$\trans$:
	\begin{align*}
		\trans = & \big\{(\state_\init, r_1), (\state_\init, t_1)\} \; \cup \; \big\{(v^j_i, v^j_{i+1}) \mid i \in [1, n) 
		\, \wedge \, j \in [1, m] \big\} \; 
		\cup\\
		& \big\{(s_i^j, {s}_{i+1}^j), (\bar{s}_i^j, {s}_{i+1}^j), ({s}_i^j, \bar{s}_{i+1}^j), (\bar{s}_i^j, 
		\bar{s}_{i+1}^j) \mid i \in [1, n) \wedge j \in [1,m] \big\} \; \cup \\
		& \big\{(s_n^j, t_{j+1}), (\bar{s}_n^j, {t}_{i+1}), (t_{j}, {s}_1^{j}), (t_{j}, 
	\bar{s}_{1}^{j}) \mid i \in [1, n) \wedge j \in [1,m] \big\} \; \cup \\
	& \big\{(v_j^n, r_{j+1}) \mid i \in [1, n] \wedge j \in [1,m) \big\} \; \cup \\
		& \big\{(t_{m+1}, t_{m+1}), (r_{m+1}, r_{m+1})\big\}.
	\end{align*}
	
\end{itemize}


\pagebreak
\noindent \textbf{A-HLTL formula: } The \AHLTL formula in our mapping is 
the following:
{
\begin{tikzpicture}
	\node (a1) at (-2.8, 1) {$\mathcal{Q}\pi'$};
	\draw[-] (-4.3,.75) -- (-1.6,.75);
	\draw[->] (-4.3,.75) -- (-4.3, .2);
	\draw[->] (-1.6,.75) -- (-1.6,.2);
	\node (a) at (-3, 0) {$\varphi_{\map} = \quant_{1} \pi_{1}.\quant_2 \pi_2\cdots \quant_n \pi_n 
		.\Etau\tau.\Etau \tau'.$};
\draw [decorate, decoration = {calligraphic brace}, thick] (4.5,-4.3) --  (4.5,-6.5);
\node (b) at (5, -5.4) {\footnotesize $\psi_3$};
\node (b) at (-.5, -4.2) {
$
\bigwedge_{d \in [1,n]}
\begin{pmatrix}
	\underbrace{\F(\start_{\pi_d, \tau} \, \wedge \, \start_{\pi', \tau} \, \wedge \, \start_{\pi_d, 
	\tau'} \, \wedge \, \start_{\pi', \tau'})}_{\psi_1} \\
	 \Until \; \Bigg[\\
	\underbrace{(\start_{\pi_d, \tau} \, \wedge \, \start_{\pi', \tau} \, \wedge \, \neg \start_{\pi_d, \tau'} \, 
	\wedge \, \neg \start_{\pi', \tau'})}_{\psi_2} \\
	 \Until \; \bigg(\\
	\Big(\start_{\pi_d, \tau} \, \wedge \, \start_{\pi', \tau} \, \wedge \, \sync_{\pi_d,\tau'} \, \wedge  \, 
	\sync_{\pi',\tau'} \; \wedge \\
	(c_{\pi_d, \tau} \leftrightarrow c_{\pi_d, \tau'}) \wedge (c_{\pi', \tau} \leftrightarrow 
	c_{\pi', \tau'}) \, \wedge \\
	 (c_{\pi_d, \tau'} \leftrightarrow c_{\pi', \tau'}) \wedge (c_{\pi_d, \tau}  \leftrightarrow c_{\pi', \tau}) 
	 \, \wedge \\
	(p_{\pi_d, \tau} \leftrightarrow p_{\pi_d, \tau'}) \, \wedge (\bar{p}_{\pi_d, \tau} \leftrightarrow 
	\bar{p}_{\pi_d, \tau'})\Big)\\
	\Until\\
	\underbrace{(\myend_{\pi_d, \tau'} \wedge \myend_{\pi', \tau'})}_{\psi_4} \bigg)\Bigg]

\end{pmatrix}
$};
\end{tikzpicture}

$$
\bigwedge
$$ 

\begin{tikzpicture}
	\draw [decorate, decoration = {calligraphic brace}, thick] (8.5, 1.6) --  (8.5,-1.8);
	\node (b) at (9, 0) {\footnotesize $\psi_5$};
	\node (b) at (3.5, 0) {
	~~~~~~	$
\Always
\begin{pmatrix}
	\sync_{\pi', \tau'} \rightarrow \Bigg[ \sync_{\pi', \tau} \, \Until \, \bigg((\neg \sync_{\pi', 
		\tau} \, \wedge \, \neg \myend_{\pi', \tau'}) \; \Until \;\\
	\bigvee_{d \in [1,n]} \Big(\big(q^{d}_{\pi_d, \tau'} 
	\leftrightarrow q_{\pi', \tau'}^d\big) \, \wedge \\
	\big((p_{\pi', \tau'} \wedge p_{\pi_d, \tau'}) \; \vee \; (\bar{p}_{\pi', \tau'} \wedge \bar{p}_{\pi_d, 
		\tau'})\big)\Big)\bigg)\Bigg] 
\end{pmatrix}
$
};
\end{tikzpicture}

Similar to the proof of Theorem~\ref{thm:alternating-single} we add the quantifier on $\pi'$ so that 
the number of alternations in $\quant_{1} \pi_{1}.\quant_2 \pi_2\cdots \quant_n 
\pi_n .\Etau\tau.\Etau \tau'$ does not change. Again, note that since $\pi'$ can only be instantiated 
with one path, namely, $\state_{\init} \cdots r_m$ the choice of the quantifier does not matter.
Intuitively, this formula expresses the following. The structure for the $\pi_d$ traces (see 
Fig.~\ref{fig:multitraj}) is for choosing truth values $\fals$ and $\tru$.
Unlike in the proof of Theorem~\ref{thm:alternating-single}, the structure is now repeated for each 
clause.
In principle, this allows for different choices in each clause; however, the \AHLTL formula 
ensures that the choices are consistent across all clauses.
For this purpose, the \AHLTL formula uses the two existential trajectories $\Etau \tau$ and $\Etau 
\tau'$.
Trajectory $\tau'$ steps through the $\pi_d$ traces and the $\pi'$ trace (representing the clauses) in 
lock step.
Trajectory $\tau$ trails behind by exactly one clause.
Let us explain the purpose of each sub-formula next:

\begin{itemize}
	\item Sub-formula $\psi_1$ ensures that the both trajectories $\tau$ and $\tau'$ are initially 
	positioned in states labeled by $\start$.
	
	\item Sub-formula $\psi_2$ holds trajectory $\tau$ in states $t_1$ and $r_1$ until trajectory 
	$\tau'$ advances to states $t_2$ and $r_2$.
	
	\item Once $\psi_2$ holds, sub-formula $\psi_3$ forces both trajectories to move in lock-step 
until trajectory $\tau'$ reaches $\myend$ states; i.e., until sub-formula $\psi_4$ holds. Sub-formula 
$\psi_3$ also ensures that the values of $p$ and $\bar{p}$ are chosen consistently between 
$\tau$ and $\tau'$. 

\item Sub-formula $\psi_5$ requires trajectory $\tau'$ to make all clauses true. Notice that $\tau'$ 
visits all states ($\tau$ does not visit states of the last clause).

\end{itemize}

We now show that the given quantified Boolean formula is $\mathit{true}$ if and 
only if the Kripke structure obtained by our mapping satisfies the \AHLTL
formula $\varphi_\map$. 

\begin{description}
	
	\item[($\Rightarrow$)] Suppose that $y$ is true. Then, there is an instantiation 
	of existentially quantified trace variables for each value of universally quantified 
	variables and the trajectory quantifier, such that each clause $y_j$, where $j \in [1, m]$ 
	becomes true (see Figs.~\ref{fig:qbf} and~\ref{fig:multitraj} for an example). We now use these 
	instantiations to instantiate each $\exists \pi_{x_d}$ in \AHLTL formula $\varphi_\map$, where 
	$d \in \{i \mid \quant_i = \exists\}$ as well as trajectories $\tau$ and $\tau'$ as follows.
	First, as mentioned earlier, $\pi'$ can only be instantiated by the trace that reaches state 
	$r_{m+1}$. Now, for each existentially quantified variable $x_i$, where $i \in [1, n]$, in depth $d \in 
	[1,k+1]$, if $x_i = \tru$, we instantiate  $\pi_d$ with a trace that includes state $s^j_i$ for the 
	clause $y_j$ that contains literal $x_i$. 
	Otherwise, the trace will include state $\bar{s}^j_i$. We now show that this 
	trace instantiation evaluates formula $\varphi_\map$ to true. 
	The trajectories can also be instantiated so that they synchronize the evaluation and consistency 
	of truth values.
	Since each $y_j$ is true, for any instantiation of universal quantifiers, there is at 
	least one literal in $y_j$ that is true. If this literal is of the form $x_i$, 
	then we have $x_i = \tru$ and trace $\pi_d$ will include $s^j_i$, which is 
	labeled by $p$ and $q^d$. Hence, the values of $p$ (respectively, $q^d$), in 
	both $\pi_d$ and $\pi'$ instantiated by trace
	$$
	\state_{\init}r_1 \cdots r_jv^j_1\cdots v^j_n\cdots r_{m+1}^\omega
	$$
	are eventually equal. This is ensured by the instantiated trajectory $\tau$.
	Similarly, if the literal in $y_j$ is of the form $\neg x_i$, 
	then $x_i = \fals$ and, hence, some trace $\pi_d$ will include $\bar{s}_i$. 
	Again, the values of $\bar{p}$ (respectively, $q^d$), in both $\pi_d$ and 
	$\pi'$ are eventually equal. Finally, since all clauses are true, all traces 
	$\pi'$ reach a state where the right side of the implication becomes true.
	
	\item[($\Leftarrow$)] Suppose our mapped Kripke structure satisfies the
	\AHLTL formula $\varphi_\map$. This means that for each instantiation of the trajectories 
	$\tau$ and $\tau$, 
	since trace $\pi'$ is of the form \linebreak $\state_{\init}r_1 \cdots r_jv^j_1\cdots v^j_n\cdots 
	r_{m+1}^\omega$, then there exists a state $v_i^j$, where the values of $q^d$ and either 
	$p$ or $\bar{p}$ are eventually equal to their counterparts in some trace $\pi_d$. This of course 
	happens in some $j$ gadget. If 
	this trace is existentially quantified and includes $s_i^j$, then we assign $x_i 
	= \tru$ for the preceding quantifications. If the trace includes $\bar{s}_i^j$, 
	then $x_i = \fals$. Observe that since in no state $p$ and $\bar{p}$ are 
	simultaneously true and no trace includes both $s_i$ and $\bar{s}_i$, variable 
	$x_i$ will have only one truth value. This is further ensured by the existence of trajectories $\tau$ 
	and $\tau'$ that guarantee the consistency of truth values.
	
\end{description}

The argument to establish \comp{${\Pi^p_{k+1}}$-} and \comp{${\Pi^p_{k+1}}$-hardness} is similar 
to that of the proof of Theorem~\ref{thm:alternating-single}.
\qed

\ \\

\subsection*{Proof of Theorem~\ref{thm:alternating-multiple-alternating}}

\newcommand\starte{\mathsf{start}^2}

First, observe that \MC{$\exists \Etau \Atau$}{} is \comp{NP-complete}. The upper bound is a trivial 
consequence of Theorem~\ref{thm:altfree-single}. The upper is also a trivial consequence of 
Theorem~\ref{thm:alternating-multiple}. Likewise, \MC{$\forall \Atau\Etau$}{} is 
\comp{coNP-complete}.

In the following, we show that the base case \MC{$\forall^+ \Etau^+\Atau^+$}{} is 
\comp{${\Pi^p_{2}}$-complete} and, by duality, \MC{$\exists^+ \Atau^+\Etau^+$}
is  \comp{${\Sigma^p_{2}}$-complete}.
The complexity for formulas with additional path quantifiers then follows analogously to the previous 
theorems.
The upper bounds for  \MC{$\forall^+ \Etau^+\Atau^+$}{} and \MC{$\exists^+ \Atau^+\Etau^+$}{} 
follow from Theorem~\ref{thm:qbf-mc-unroll-new} (i.e., the soundness of our BMC algorithm for 
formulas with one alternation for trajectory quantifiers). 

For the lower bound, we encode the satisfiability of a QBF 
formula \linebreak $\quant_1 x_1.\quant_2 x_2\dots\quant_{n-1} 
x_{n-1}.\quant_n x_n.(y_1 \, \wedge \, y_2 \, 
\wedge \dots \wedge \, y_m)$ with a {\em single} quantifier alternation such that $\quant_1 = 
\quant_2= \ldots = \quant_k = \forall$ and $\quant_k = \quant_{k+1} = \ldots = \quant_n = \exists$ as 
a model checking problem of a $\forall^+ \Etau^+\Atau^+$ \AHLTL formula.
We choose a single alternation for simplicity. More alternations will follow in the same fashion as 
Theorems~\ref{thm:alternating-single} and~\ref{thm:alternating-multiple}.

We modify the Kripke structure from the proof of 
Theorem~\ref{thm:alternating-single} by adding a fresh path that starts 
with a state labeled $\{\starte\}$ and then alternates for $n-k$ times 
between states labeled $\{p\}$ and states labeled $\{\bar{p}\}\}$.
The role of this path is to encode the values for the existentially chosen 
variables $x_{k+1},\ldots, x_n$. By aligning $\{p\}$ and $\{\bar{p}\}\}$ 
positions with the positions of the paths representing the truth assignments 
of the preceding quantifiers, the trajectory quantifier picks a truth 
assignment for the innermost quantifiers.
This is reflected in the new \AHLTL formula $\varphi_{\map}$:\\

\newpage

$\varphi_{\map} = \forall \pi_a \forall \pi_e \forall \pi'. \Etau \tau 
.\Atau \tau'. $ \\
$$
\underbrace{\bigg( \F \start_{\pi_a,\tau} \wedge \F \starte_{\pi_e,\tau} \, 
\wedge \, \F \sync_{\pi',\tau}\bigg)}_{\psi_1}
$$
$$\rightarrow$$
\vspace{-3mm}
\begin{tikzpicture}
\draw [decorate, decoration = {calligraphic brace}, thick] (-5,0) --  
(-5,2.5);
\node (b) at (-5.5, 1.2) {\small $\psi_3$};
	\node (a) at (-1, 0) {
$
\begin{pmatrix} 
\underbrace{\Big((\neg \start_{\pi_a,\tau} \wedge \neg \starte_{\pi_e,\tau}) 
\Until (\Box \sync_{\pi', \tau})\Big)}_{\psi_2} \; \wedge \\
\Always \Big(\neg \start_{\pi_a,\tau} \wedge c_{\pi_a,\tau} \wedge 
p_{\pi_e,\tau} \rightarrow (p_{\pi_e,\tau} \Until \neg{c}_{\pi_a,\tau})\Big) \; 
\wedge\\
 \Always \Big(\neg \start_{\pi_a,\tau} \wedge \neg{c}_{\pi_a,\tau} \wedge 
 p_{\pi_e,\tau} \rightarrow  (p_{\pi_e,\tau} \Until c_{\pi_a,\tau})\Big) \; 
 \wedge \\
\Always \Big(\neg \start_{\pi_a,\tau} \wedge c_{\pi_a,\tau} \wedge 
\bar{p}_{\pi_e,\tau} \rightarrow  (\bar{p}_{\pi_e,\tau} \Until 
\neg{c}_{\pi_a,\tau})\Big) \; \wedge \\
\Always \Big(\neg \start_{\pi_a,\tau} \wedge \neg{c}_{\pi_a,\tau} \wedge 
\bar{p}_{\pi_e,\tau} \rightarrow  (\bar{p}_{\pi_e,\tau} \Until 
c_{\pi_a,\tau})\Big) \; \wedge\\
	\Always\Bigg[ \underbrace{\Big(\sync_{\pi',\tau'}  \wedge \big(\sync_{\pi', \tau'} \Until (\neg 
		\sync_{\pi', \tau'} 	\wedge 	(c_{\pi_a,\tau} \Iff c_{\pi',\tau'})) \Until 
		\sync_{\pi',\tau'}\big)\Big)}_{\psi_4}\\
	\rightarrow \\
	 \F \Big(\big(q^{1}_{\pi',\tau'} \wedge \big((p_{\pi',\tau'} \leftrightarrow 
	 p_{\pi_e,\tau}) 
	 \vee (\bar{p}_{\pi',\tau'} \wedge \bar{p}_{\pi_e,\tau})\big)\big) \; \vee\\
\big(q^{2}_{\pi',\tau'} \wedge \big((p_{\pi',\tau'} \leftrightarrow p_{\pi_a,\tau}) \vee (\bar{p}_{\pi',\tau'} 
\leftrightarrow \bar{p}_{\pi_a,\tau})\big)\big)\Big)	 \Bigg] 
\end{pmatrix}
$
	};
\draw [decorate, decoration = {calligraphic brace}, thick] (-5,-3.8) --  
(-5,-1.9);
\node (c) at (-5.5, -2.7) {\small $\psi_5$};

\end{tikzpicture}

\ \\

Trace $\pi_a$ holds the valuation of the universal variables, trace $\pi_e$ 
together with trajectory $\tau$ the valuation of the existential variables.
Trajectory $\tau$ aligns the clauses with the valuation (analogously to 
$\tau$ in the proof of Theorem~\ref{thm:alternating-single}).
The intended purpose of the sub-formulas are as follows:

\begin{itemize}
	\item Sub-formula $\psi_1$ initialize the paths at the right 
	initial place to go through the clauses and propositional variables.
	
	\item The role of sub-formula $\psi_2$ is to create enough ``slack'' in the
	trajectory $\tau$ so that $\tau'$ can align the variables with every 
	possible clause. For this purpose, $\tau$ waits until $\pi'$ has reached 
	the terminal state before advancing $\pi_a$ and $\pi_e$.

	\item Sub-formula $\psi_3$ ensures that at all times, traces $\pi_a$ and 
	$\pi_e$ toggle between $c$ and $\neg c$. Furthermore, the four 
	conjuncts ensure that trajectory $\tau$ only assigns a single valuation to 
	each variable.
	
	\item Sub-formulas $\psi_4$ and $\psi_5$ have exactly the same role as 
	in the mapping in Theorem~\ref{thm:alternating-single}: paths $\pi_a$ 
	(and $\pi_e$) and $\pi'$ advance in lock step in trajectories $\tau$ and 
	$\tau'$ to evaluate clauses. 
\end{itemize}  
  
The (if and only if) reduction are similar to that of 
Theorem~\ref{thm:alternating-single}.
The complexity for formulas with additional path quantifiers then follows 
analogously to the previous theorems.

	\section{Details of our Case Studies}
\label{sec:appendix:detail-example}

\subsection{Counterexample for the Program in Fig.~\ref{fig:concleak_2}}
\label{appendix:concurrent}

One possible interleaving is as follows.
First, \code{T3} executes lines 1--3 and set both \code{h} and \code{l} to 1. 
Next, since \code{h = l = 1}, \code{T1} runs lines 1--5, changes \code{x} to \code{1} while 
\code{y} remains 
\code{0}. 
At this moment, \code{T2} executes lines 1--3, which gives a sequence of outputs  \{ 
\code{1}, 
\code{0} \}. 
Then, \code{T1} resumes and changes \code{y} to \code{1}, so in the next iteration 
\code{T2} prints another sequence of outputs \{ \code{1}, \code{1} \}, and all threads halt.

This sequence of outputs  \{ \code{1}, \code{0}, \code{1}, \code{1} \} printed by this specific 
scenario 
leaks the information that \code{h = 1}, because when \code{h = 0}, \code{x} can be set to 
\code{1} 
only if \code{y} has been set to \code{1} too. Since \code{T2} always prints \code{x} before 
\code{y}, 
so the output sequence  \{ \code{1}, \code{0}, \code{1}, \code{1} \} is not reproducible; 
hence, leaks 
information 

Note that this specific information leak happens when the sequence  of public inputs can be 
observed by the 
attacker. In other words, an attacker can correctly guess the value of \code{h} by observing 
this particular 
sequence of public inputs and  outputs.  
That is, in order to correctly detect information leakage, one has to use a formula that aligns 
the public inputs 
and the observable outputs separately because the two alignments may ``cross'' with each 
other as 
illustrated in Section~\ref{sec:intro}.  
As a result, two trajectory variables are needed in this case, as presented in the formula 
$\varphi_\propODndet$ below.
%
Our encoding allows correct detection for this information leak due to concurrency (as 
shown in 
Table~\ref{tab:exp_results}). The returned counterexample presents the concurrent bug 
where the secret 
inputs can be guessed by a malicious attacker from observing the  sequences of public 
inputs and outputs.

\subsection{Detail Explanation of Speculative Information Flows}
\label{appendix:speculative}

\begin{figure}[t!]
	\centering
	\input{figs/progSE}
	\caption{The speculative execution example, where the secure translation is applied load 
	hardening technique to avoid information leaks. }
	\label{fig:speculative_execution}
	\vspace{-5mm}
\end{figure}

The \code{C} program in Fig.~\ref{fig:se_ccode} simply performs array data accessing with a given 
input index  \code{y}, where the \code{if} statement checks if \code{y} stays in the ideal bound of 
array size or not.
However, since low level assembly code requires more steps on memory storing and accessing 
using registers, an attacker might be able to guess the high-security values by comparing the 
speculative and non-speculative executions after translation (see Fig.~\ref{fig:se_assembly1}). 
In this case study, we evaluate 7 different versions of code where some of them are secure and 
some of them are insecure when doing speculative runs.
%
%
\begin{itemize}
	\renewcommand{\labelitemi}{$\bullet$}
	\item $\progse_{V1}$: {\em Insecure array value accessing.}
	The first version considers naive translation as shown in Fig.~\ref{fig:se_assembly1}.
	In non-speculative cases (i.e., \code{y} $<$ \code{size}), the observations are identical.
	However, in speculative executions (i.e.,  \code{y} $\geq$ \code{size} ), lines 5 -- 8 in the 
	assembly code  (see Fig.~\ref{fig:se_assembly1}) leak the value of the critical information (i.e., line 
	2 in the original \code{C} program). Thus, it violates SNI. 
	Our BMC technique successfully returns UNSAT in this case, indicating a counterexample has 
	been spotted.

	\item $\progse_{V2}$: {\em Secure array value accessing with masking.} 
	The second version shown in Fig.~\ref{fig:se_assembly2} overcomes the leaks from 
	$\progse_{V1}$ by applying the countermeasure of {\em speculative load hardening.} By adding an 
	extra variable {\em mask} which the value is assigned on line 3 and 6, followed by the two 
	\code{or} operations on lines 9 and 11, {\em mask} successfully hides the real value in a 
	speculative execution which an attacker might try to access. Hence, it satisfies SNI. Our BMC 
	algorithm returns SAT and shows the absence of a counterexample.
	
\end{itemize}

\noindent{\bf $\progse_{V1}$: Insecure array value accessing}

We first consider naive translation as showed in Fig.~\ref{fig:se_assembly1}.
In non-speculative cases (i.e., \code{y} $<$ \code{size}), the observations of executions are identical. However, when for speculative executions (i.e.,  \code{y} $\geq$ \code{size} ), line 5 -- 8 will leak value of the critical information (i.e., line 2 in the c program).
Thus it violates SNI and our solver successfully returns SAT, indicating a counterexample.

\vspace{2mm}
\noindent{\bf $\progse_{V2}$: Secure array value accessing with masking}

The second version showed in Fig.~\ref{fig:se_assembly2} overcomes the leaks from SE$_{V1}$ by apply the countermeasure of {\em speculative load hardening.} 
By adding an extra variable {\em mask} which the value is assigned on line 3 and 6, where the \code{or} operations on line 9 and 11 hides the real value that a speculative execution might try to access. Hence, it satisfies SNI, where our solver returns UNSAT.

\vspace{2mm}
\noindent{\bf $\progse_{V3}$: Speculative Load Hardening (insecure) }

The third version refines $V_1$ with another if-statement.
However, even with this extra conditional checking, the translation still violates speculative non-interference because the 
attacker can now know secret information about the content of \code{array A} in index \code{y}. (i.e., If it is equivalent to \code{k}) or not.  

\vspace{2mm}
\noindent{\bf $\progse_{V4}$: Branching (insecure) }

The fourth version is investigating the usage of conditional operator. 
The compilation here translates the conditional operator directly into a 
branch instruction. 
This additional branching creates extra source for speculative execution and hence, 
leads to a potential speculative leak.

\ \\

\vspace{2mm}
\noindent{\bf $\progse_{V5}$: Conditional Move (secure) }

The fifth version conquers the problem in $\progse_{V4}$. 
In this case, the conditional operators are always translated into conditional moves, 
instead of creating harmful branching. 
As a result, $\progse_{V5}$ is secure under the speculative runs.

\vspace{2mm}
\noindent{\bf $\progse_{V6}$: Pointer (insecure) }

The last two versions $\progse_{V6}$ and $\progse_{V7}$ are considering how the input pointer provided by an attacker could cause speculative leaks. 
Essentially, $\progse_{V6}$ assume the attacker specify the input as a pointer value, the memory access in the speculative run would be exploited by the attacker due to secondary memory access. 
In other words, the attacker is able to obtain the sensitive information by assigning a pointer input. As a result, insecure. 

\vspace{2mm}
\noindent{\bf $\progse_{V7}$: Pointer with Load Hardening (secure) }

On the other hand, $\progse_{V7}$ resolves this problem by performing hardening. 
In this way, no harmful information flow would happen in the speculative runs. Hence, 
the program is proved secure.

%

\subsection{Secure Compiler Optimization}
\label{appendix:compileropt}

In Fig.~\ref{fig:LP}, if the outermost while-loop executes only one time, the  
alignment of public outputs requires only one existential trajectory (i.e., $\varphi_\opt$).
Figure~\ref{fig:LP} peels off the first iteration of a for-loop (LP).

However, when the outermost while loop execute several times and the inputs are read from 
asynchronous channels, two trajectories are required to check conformance of source and target 
(i.e., $\varphi_\optndet$).
We also notice that our trajectory encoding is able to reproduce the same verification outcomes as 
introduced in case studies in~\cite{bcbfs21} with better performance in time

\begin{figure}[t]
	\input{figs/progCO}
	\caption{A compiler optimization example of loop peeling with nondeterministic input sequence.}
	\label{fig:LP}
\end{figure}

\begin{figure}[t]
		\input{figs/progCacheTA}
	\caption{Cache-based timing attack with nondeterministic inputs from 
		thread T4. }
	\label{fig:CacheTA}
		\vspace{-2mm}
\end{figure}

\subsection{Cache-based Timing Attacks}
\label{appendix:chcheattack}

We consider a 4-threaded program as presented in Fig.~\ref{fig:CacheTA}  
The three threads \code{T1}, \code{T2}, and \code{T3} are interacting with 
the cache when they 
execute \code{fillArray()} or \code{readArray()}.
The counters $n$, $m$, and $p$ are representing the numbers of 
execution steps for threads 
\code{T1}, \code{T2}, and \code{T3}, respectively, to schedule the threads 
in the order of \code{T1}, 
\code{T2}, and \code{T3}. 

Assume the cache size is $M=2$ and  is originally filled with the value of 
the low array.
When \code{secret} does not equal to \code{low}, the cache will remain 
with low array data (i.e., 
\code{T1} doesn't execute line 3--4).
Hence, when \code{T2} starts executing line 4, it only needs to read the 
data which takes two steps 
(since $M=2$), and then outputs 1.
Afterward, \code{T3} reads the data in line 4 again, and outputs 0, which 
leads to the final sequence 
of outputs $\{1, 0\}$. Now, consider 
\begin{wrapfigure}{r}{.35\textwidth}
\centering
	\includegraphics[scale=0.8]{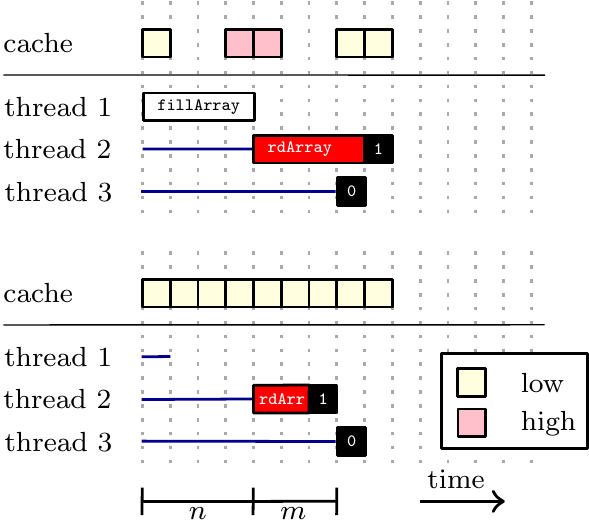}
	\caption{A Cache-based timing attack. }
	\label{fig:demo_CacheTA}
\end{wrapfigure}
when \code{secret} equals to \code{low}, which makes the cache filled with 
high-array data (i.e., \code{T1} executes line 3--4). 
In this case, \code{T2} will take longer time to finish line 4, because of the 
necessary operations of 
evicting the high-array data, which adds extra steps for \code{T2} to finish 
line 4 compare to the 
previous scenario. 
Hence, \code{T3} will start before \code{T2} finished, and output 0 before 
\code{T2} eventually 
output 1. 
That is, an output sequence $\{0, 1\}$. The two scenarios are presented in 
Fig.~\ref{fig:demo_CacheTA}

We modify this original example in~\cite{stefan2013eliminating} with 
another source of 
nondeterminism, the input channel \code{T4}, to show that two trajectories 
are needed in this case
in order to correctly identify a cache-based timing attack. 
With \code{T4}, one has to align the inputs first before aligning the outputs 
from \code{T2} and 
\code{T3} when there are sequences of inputs and outputs (due to the 
loops).

\end{document}